\documentclass[a4paper,11pt,amsmath,amssymb]{article} 
\usepackage{jheppub}
\usepackage{amsmath} 
\usepackage{array,relsize,float}
\usepackage{pstricks}
\usepackage{color} 
\usepackage{amssymb}
\usepackage{slashed}
\usepackage{tikz-cd}
\usepackage{graphicx} 
\usepackage{epsfig}
\usepackage{multicol} 
\usepackage{slashed}
\usepackage{tensor}
\usepackage{hyperref}
\usepackage{pdfpages}
\usepackage{array}
\usepackage{tikz-cd}
\usepackage{amsfonts,layout,appendix,subfigure}
\allowdisplaybreaks
\input{epsf}

\renewcommand{\thefigure}{\arabic{figure}}

\def\beq{\begin{equation}} \def\eeq{\end{equation}}
\def\beqn{\begin{eqnarray}} \def\eeqn{\end{eqnarray}}
 \def\to{\rightarrow}
\def\nn{\nonumber}

\def\ln#1{\mathrm{log}\left(#1\right)}

\def\beq{\begin{equation}}
\def\eeq{\end{equation}}
\def\bea{\begin{eqnarray}}
\def\eea{\end{eqnarray}}
\def\beqn{\begin{eqnarray}} \def\eeqn{\end{eqnarray}}
\def\beeq{\begin{eqnarray}}
\def\eeeq{\end{eqnarray}}

\def\nn{\nonumber}

\def\ln#1{\mathrm{log}\left(#1\right)}

\newcommand{\qcdqed}{QCD$\otimes$QED}
\usepackage{color,xcolor,soul}

\def\qt{$q_T$} 

\newcommand\as{\alpha_{\mathrm{S}}}

\def\as{\alpha_{\rm S}}

\def\muRsq{\mu_R^2}
\def\muFsq{\mu_F^2}

\newcommand{\valencia}{Instituto de F\'{\i}sica Corpuscular (IFIC), CSIC-Universitat de Val\`{e}ncia, Parc Cient\'{\i}fic, E-46980 Paterna, Valencia, Spain.}
\newcommand{\cern}{CERN, CH-1211 Geneva, Switzerland.}

\newcommand{\milano}{Tif Lab, Dipartimento di Fisica, Universit\'a di Milano and INFN, Sezione di Milano, Via Celoria 16, I-20133 Milan, Italy.}
\newcommand{\salamanca}{Departamento de F\'isica Fundamental e IUFFyM, Universidad de Salamanca, E-37008 Salamanca, Spain.}

\begin{document}

\title{Transverse-momentum resummation at mixed \qcdqed~NNLL accuracy for \boldmath$Z$ boson production at hadron colliders}

\author[a]{Andrea Autieri,}
\author[b]{Stefano Camarda,}
\author[a]{Leandro Cieri,}
\author[c]{Giancarlo Ferrera}
\author[d]{and German~F. R.~Sborlini} 
\affiliation[a]{\valencia}
\affiliation[b]{\cern}
\affiliation[c]{\milano}
\affiliation[d]{\salamanca}

\emailAdd{andrea.autieri@ific.uv.es}
\emailAdd{stefano.camarda@cern.ch}
\emailAdd{leandro.cieri@ific.uv.es}
\emailAdd{giancarlo.ferrera@mi.infn.it}
\emailAdd{german.sborlini@usal.es}

\preprint{FTUV-25-0731.8337}

\abstract{We consider the transverse momentum (\qt) distribution of neutral charged bosons at hadron colliders. We perform the resummation of the logarithmically-enhanced effects due to simultaneous QCD and QED initial-state radiation, up to mixed next-to-next-to-leading logarithmic (NNLL) accuracy. We study the impact of such mixed \qcdqed~resummed contributions on top of pure QCD corrections, finding percent-level effects.}

\setcounter{page}{1}
\maketitle

\section{Introduction}
\label{sec:introduction}
The Large Hadron Collider (LHC) \cite{Evans:2008zzb} at CERN is reaching an impressive level of precision for many relevant processes in high-energy physics. Among these processes, a central role is played by the Drell-Yan (DY) mechanism
\cite{Drell:1970wh,Christenson:1970um}, which consists in the production of 
high-invariant mass lepton pairs through the decay of an electroweak (EW) boson ($\gamma^{*}$, $Z$, $W$). The presence of a clear experimental signature with at least one charged lepton in the final state and its high production rate, makes DY very relevant for detector calibration, luminosity monitor and also for probing the underlying event. Furthermore, it is useful for extraction of Standard Model parameters, such as parton density functions \cite{Accomando:2017scx}, the Weinberg angle \cite{CMS:2024ony,CMS:2011utm}, the $W$ boson mass \cite{CDF:2013dpa,ATLAS:2017rzl,CDF:2022hxs,ATLAS:2023fsi} and its decay width \cite{Camarda:2016twt}, and the strong coupling \cite{Camarda:2022qdg,ATLAS:2023lhg}. 

Among the various kinematic distributions, the transverse momentum ($q_T$) distribution plays a significant role. In particular, the analysis of the $Z$ boson spectra provides us important information
about the mechanism of $W$ boson production. Moreover, a precise knowledge of the $W$ boson transverse momentum distribution at intermediate and large \qt~is important for the determination of its mass \cite{Rottoli:2023xdc,CDF:2013dpa,ATLAS:2017rzl,CDF:2022hxs,ATLAS:2023fsi,CarloniCalame:2016ouw}. Regarding the $Z$ boson transverse momentum distribution, it is relevant for precise measurements of the strong coupling \cite{Camarda:2022qdg,ATLAS:2023lhg}. 

In the region where the transverse momentum is of the order of the vector boson mass $M$, the fixed-order approximations are reliable. In this region the \qt-distribution is known up to $\mathcal{O}(\as^3)$ through numerical QCD calculations of V+jet production at NNLO \cite{Boughezal:2015dva,Boughezal:2015ded,Gehrmann-DeRidder:2016cdi,Gehrmann-DeRidder:2017mvr,Chen:2022cgv,Alioli:2025hpa}.

Given the high level of experimental precision achieved in recent years \cite{CMS:2019raw, ATLAS:2025hhn,ATLAS:2024nrd, ATLAS:2024irg,CMS:2024myi, LHCb:2016fbk}, providing accurate theoretical predictions for the DY $q_T$ spectrum has become essential for a reliable phenomenological description. Along this line, it is essential to include higher-order radiative corrections in QCD and simultaneously, in QED, since the two couplings are roughly related by $\as^2 \sim \alpha$. In mixed strong-electroweak theory, contributions to the differential distributions are known at $\mathcal{O} (\alpha)$ \cite{CarloniCalame:2006zq,CarloniCalame:2007cd}, at $\mathcal{O} (\as \alpha )$ for neutral current DY \cite{Bonciani:2021zzf,Armadillo:2022bgm,Armadillo:2022ugh,Buccioni:2022kgy} and charged current DY, neglecting the exact NNLO virtual contribution \cite{Buonocore:2021rxx}.

When we consider the kinematic region $q_T \ll M$, large logarithmic terms 
spoil the convergence of the fixed-order calculation. These terms must be evaluated and resummed at all orders to restore the predictability of the perturbation theory. The procedure for systematically performing the resummation is known since a long time \cite{Parisi:1979se,Kodaira:1981nh,Altarelli:1984pt,Collins:1984kg,Catani:2000vq}. Nowadays, different resummation formalisms have been developed in order to reach the next-to-next-to-next-to-leading logarithm ($\text{N}^3\text{LL}$) accuracy \cite{Camarda:2021ict,Bizon:2018foh,Bizon:2019zgf,Bacchetta:2019sam,Ebert:2020dfc,Becher:2020ugp,Re:2021con,Ju:2021lah,Campbell:2022uzw,Campbell:2023cha,Isaacson:2023iui, Bacchetta:2022awv,Bacchetta:2024qre,Bacchetta:2025ara,Camarda:2025lbt}, and the following order, i.e. $\text{N}^4\text{LL}$ \cite{Neumann:2022lft,Camarda:2023dqn,Moos:2025sal,Billis:2024dqq,Piloneta:2024aac}.  


As for the fixed-order case, the resummation of QED logarithmic terms is important in order to produce precise results. Mixed \qcdqed~resummation at NLL+NLO was performed for on-shell $Z$ boson production \cite{Cieri:2018sfk} and on-shell $W$ boson production \cite{Autieri:2023xme}. The off-shell case at NLL+NLO was considered in Ref. \cite{Buonocore:2024xmy}. In this article, we go beyond the current precision frontier by considering, for the first time, the resummation of mixed \qcdqed~contributions at NNLL accuracy for the neutral current DY process. Our formalism takes into account initial-state radiation (ISR) and is valid for the production of any collection of chargeless particles in the final state 

The outline of this paper is the following. In Sec. \ref{sec:QCDresummation}, we review the QCD transverse-momentum resummation formalism. Then, we describe the simultaneous and consistent combination of QED and QCD resummation in Sec. \ref{sec:CombinedQEDQCD}. There, we explicitly present all the ingredients required to reach up to NNLL mixed \qcdqed~accuracy. The results for $Z$ and $\gamma^*$ production are described in Sec. \ref{sec:ResultsZ}, considering Tevatron and LHC kinematics. Finally, we discuss the conclusions and future research directions in Sec. \ref{sec:Conclusions}.


\section{QCD transverse-momentum resummation formalism}
\label{sec:QCDresummation}
In this section, we address the transverse-momentum resummation formalism of Refs. \cite{Catani:2000vq,Bozzi:2005wk} for vector boson production at hadron colliders. We consider the following process:
\begin{equation}
    h_1 (p_1) + h_2(p_2) \rightarrow V(M, q_T) + X
\end{equation}
where $p_1$ and $p_2$ are the momenta of the colliding hadrons, $M$ and $q_T$ are the invariant mass and the transverse momentum of the triggered final state, respectively. Here, $X$ denotes undetected radiation under which we are totally inclusive. 

According to the factorization theorem \cite{Collins:1989gx}, the $q_T$-differential cross-section can be written as: 
\beqn
\nn \frac{d \sigma_{h_1 h_2 \to V}}{dq_T^2}(q_T,M,s) &=& \sum_{a_1, a_2} \int dx_1 dx_2 \, f_{a_1/h_1}(x_1,\mu_F^2) \,  f_{a_2/h_2}(x_2,\mu_F^2) \, 
\\ &\times& \frac{d \hat{\sigma}_{a_1 a_2 \to V}}{dq_T^2}(q_T,M,\hat{s};\muFsq) \, ,
\label{eq:Master1}
\eeqn
where $\sqrt{s}$ is the hadronic centre-of-mass energy ($s = (p_1+p_2)^2 \simeq 2 p_1 p_2$), $f_{a_i/h_i} (x_i, \mu_F^2)$ are the parton distribution functions (PDF) at the factorization scale $\mu_F$, $\hat{\sigma}_{a_1 a_2}$ is the partonic cross section, $\hat{s} = x_1 x_2 s$ is the partonic centre-of-mass energy and $\mu_R$ is the renormalization scale. The partonic cross-section can thus be calculated using perturbation theory, whereas the non-perturbative dependence of the hadron process is universal and embodied within the PDFs. 

Higher-order perturbative contributions to the partonic cross-section are enhanced in the small $q_T$ region (i.e. $q_T \ll M $) by logarithmic terms proportional to 
\beq 
(\as^n/q_T^2) \log^m ( M^2/q_T^2) \, . \eeq
Therefore, we explicitly split the partonic cross-section as follows: 
\begin{equation}
\label{Form_Res_Fin}
    \frac{d \hat{\sigma}_{a_1 a_2 \rightarrow V}}{d q_T^2} = \frac{d \hat{\sigma}^{\text{res.}}_{a_1 a_2 \rightarrow V}}{d q_T^2} + \frac{d \hat{\sigma}^{\text{fin.}}_{a_1 a_2 \rightarrow V}}{d q_T^2},
\end{equation}
where $d \sigma^{\text{res.}}$ contains the large logarithmic terms while the finite remainder $d \hat{\sigma}^{\text{fin.}}$ fulfills
\begin{equation}
    \lim_{Q_T \rightarrow 0} \int_0^{Q_T^2} d q_T^2 \frac{d \hat{\sigma}^{\text{fin.}}}{d q_T^2} = 0 \ ,
\end{equation}
and is, by definition, free of logarithmically-enhanced contributions. In this article, we deal only with the resummed part, which is dominant in the small $q_T$-region \cite{Bozzi:2010xn}.

In order to impose transverse-momentum conservation, the resummation is performed in impact parameter space $b$, which is the Fourier-Bessel conjugated to $q_T$. Thus, we have
\begin{equation*}
\frac{d \hat{\sigma}_{a_1 a_2 \rightarrow V}^{\rm{res.}}}{d q_T^2} (q_T; M, \hat{s}; \as (\mu_R^2), \mu_F^2, \mu_R^2) 
= \frac{M^2}{\hat{s}} \int_0^{\infty} db \frac{b}{2} J_0(bq_T)
\end{equation*}
\begin{equation}
\label{Bessel-res}
\times \mathcal{W}_{a_1 a_2}^V (b; M, \hat{s}, \as (\mu_R^2), \mu_F^2, \mu_R^2),
\end{equation}
with $J_0(x)$ the 0th-order Bessel function. Then, we consider the Mellin $N$-moments with respect to $z = M^2/\hat{s}$ at fixed $M$:
\begin{equation}
    \mathcal{W}^V_{ab,N} (b,M; \as(\mu_R^2),\mu_R^2,\mu_F^2) = \int_0^1 dz \, z^{N-1} \mathcal{W}^V_{ab}(b,M, \hat{s} = M^2/z; \as(\mu_R^2), \mu_R^2, \mu_F^2 ).
\end{equation}
In the simplified flavour diagonal case $(a,b) = (q,\bar{q})$ in the Mellin space, we have an exponentiation of the logarithmic terms according to: 
\begin{equation*}
    \mathcal{W}_N^V(b,M; \as(\mu_R^2),\mu_R^2,\mu_F^2 ) = \mathcal{H}_N^V(M,\as(\mu_R^2); M^2/\mu_R^2, M^2/\mu_F^2, M^2/Q^2 )
\end{equation*}
\begin{equation}
\label{Mellin-resum}
  \times  \exp  \left\{ \mathcal{G}_N(\as(\mu_R^2),L; M^2/\mu_R^2, M^2/Q^2 ) \right\},
\end{equation}
where
\begin{equation}
    L = \log \left( \frac{Q^2 b^2}{b_0^2}  \right),
\end{equation}
$Q$ is the resummation scale (introduced to parametrize the arbitrariness of the factorization between the fixed order and resummed contributions) and $b_0 =  2 e^{- \gamma_E}$, being $\gamma_E=0.5772 \ldots$ the Euler-Mascheroni constant.

Studying the ingredients of Eq. (\ref{Mellin-resum}), $\exp\{ {\mathcal{G}}_N\}$ contains and resums all the logarithmic enhanced terms while $\mathcal{H}^{V}$ includes hard-virtual contributions that behave like constants as $b \rightarrow \infty$. The exponent $\mathcal{G}_N$ can be systematically expanded as: 
\begin{equation}
\label{G-expansion-g}
\mathcal{G}_N (\as, L; M^2/\mu_R^2, M^2/Q^2)= L \, g_N^{(1)} (\as L) + g_N^{(2)} (\as L) + \sum_{n=3}^{+\infty} \left( \frac{\as}{\pi} \right)^{n-2} g_N^{(n)} (\as L) \, .
\end{equation}
Therefore Eq. (\ref{Mellin-resum}) constitutes a perturbative expansion in the parameter $\as L$: $L g^{(1)}$ resums the \textit{leading logarithmic} (LL) contributions $\as^n L^{n+1}$, $g_N^{(2)}$ together with $\mathcal{H}_N^{V(1)}$ control the \textit{next-to-leading logarithmic} (NLL) contributions $\as^n L^{n}$, $g_N^{(3)}$ together with $\mathcal{H}_N^{V(2)}$ collect the \textit{next-to-next-to-leading logarithmic} (NNLL) terms $\as^n L^{n-1}$ and so on.

The functions $g_N^{(n)}$ can be explicitly obtained from the integral representation of $\mathcal{G}_N$:  
\begin{equation}
\label{integral_rep_G}
\mathcal{G}_N \big( \as(\mu_R^2), L; M^2/\mu_R^2, M^2/Q^2 \big) =
- \int_{b_0^2/b^2}^{Q^2} \frac{dq^2}{q^2} 
\left[ A(\as(q^2)) \log \frac{M^2}{q^2} + \widetilde{B}_N(\as(q^2)) \right]
\end{equation}
where $A(\as)$ and $\widetilde{B}_N(\as)$ are perturbative series in the strong coupling that can be obtained from the logarithmic terms in the perturbative expansion of the $q_T$ distribution at the $n$-th relative order in $\as$ \cite{Bozzi:2005wk}.

The explicit expression of the $g_N^{(1)}$, $g_N^{(2)}$ and $g_N^{(3)}$ coefficients can be found in Ref.~\cite{Bozzi:2005wk}, $g_N^{(4)}$ in Ref.~\cite{Bizon:2017rah} for the related case of direct transverse momentum space resummation and $g_N^{(5)}$ in Ref.~\cite{Camarda:2023dqn}.

The hard-virtual coefficient $\mathcal{H}_N^V$, can be expanded as:
\begin{equation}
\label{H-expansion}
    \mathcal{H}_N^V = 1  + \sum_{n=1}^{\infty} \left( \frac{\as}{\pi} \right)^n \mathcal{H}_N^{V(n)} \, ,
\end{equation}
with $\mathcal{H}^{V(n)}$ depending on the process. These coefficients can be written in terms of $H^V(\as)$ (which is associated to finite contribution of the multi-loop virtual scattering amplitude) and the process-independent functions $C(\as)$ related to the collinear emissions from the initial-state partons according to \cite{Catani:2012qa,Catani:2013tia}: 
\begin{equation}
        \mathcal{H}_{q\bar{q} \to ab}^{V}(z; \as) = H_q^{V}(\as) 
\int_0^1 dz_1 \int_0^1 dz_2 \, \delta(z - z_1 z_2) 
C_{q a}(z_1; \as) C_{\bar{q} b}(z_2; \as)\,.
\end{equation}
The hard-virtual factor can be expanded in perturbative QCD series as
\begin{equation}
H^V(\as) = 1 + \sum_{n=1}^{\infty} H^{V(n)},
\end{equation}
and the collinear coefficient functions,
\begin{equation}
    C(\as) = 1 + \sum_{n=1}^{\infty} \left( \frac{\as}{\pi} \right)^n C^{(n)}.
\end{equation}

At NLL+NLO we include the functions $g_N^{(1)}$, $g_N^{(2)}$ and ${\cal H}_N^{V(1)}$, 
at NNLL+NNLO we also include the functions $g_N^{(3)}$ and ${\cal H}_N^{V(2)}$\,\cite{Catani:2012qa,Gehrmann:2012ze}, at N$^3$LL+N$^3$LO  the functions $g_N^{(4)}$ and ${\cal H}_N^{V(3)}$\,\,\cite{Luo:2019szz,Ebert:2020yqt} and finally at N$^4$LL+N$^3$LO  the function $g_N^{(5)}$ and ${\cal H}_N^{V(4)}$.


\section{Combined QED and QCD transverse-momentum resummation formalism}
\label{sec:CombinedQEDQCD}
When simultaneously dealing with QCD and QED corrections, the resummed component in Mellin and b-space can be written as \cite{Autieri:2023xme,Cieri:2018sfk}:
\begin{equation}
    \begin{aligned}
        \mathcal{W}_{N}^{' V}(b, M; \mu_F) = 
        & \, \hat{\sigma}_{V}^{(0)}(M) \, 
        \mathcal{H}_{N}^{' V} (\as, \alpha; M^2/\mu_R^2, M^2/\mu_F^2, M^2/Q^2) \\
        & \times \exp \left\{ \mathcal{G}_{N}' (\as, \alpha, L; M^2/\mu_R^2, M^2/Q^2) \right\},
    \end{aligned}
\end{equation}
where the resummed form factor and the hard-collinear coefficient functions have a double perturbative expansion in $\as$ and $\alpha$, respectively. Thus we write:
\begin{equation}
\begin{array}{rl}
\mathcal{G}'_N (\as, \alpha, L) = & \mathcal{G}_N (\as, L) 
+ L \, g'^{(1)}(\alpha L) 
+ g'^{(2)}_N(\alpha L) 
+ \sum\limits_{n=3}^{\infty} \left( \frac{\alpha}{\pi} \right)^{n-2} g'^{(n)}_N (\alpha L) \\[10pt]
& + g'^{(1,1)}(\as L, \alpha L) 
+ \sum\limits_{\substack{n,m=1 \\ n+m \neq 2}}^{\infty}
\left( \frac{\as}{\pi} \right)^{n-1}
\left( \frac{\alpha}{\pi} \right)^{m-1}
g'^{(n,m)}_N (\as L, \alpha L) \, ,
\label{eq:gmixedcoefs}
\end{array}
\end{equation}
and
\begin{equation}
\begin{array}{rl}
\mathcal{H}'^{V}_N (\as, \alpha) = & \mathcal{H}^{V}_N (\as) 
+ \dfrac{\alpha}{\pi} \mathcal{H}^{'V(1)}_N
+ \sum\limits_{n=2}^{\infty} \left( \dfrac{\alpha}{\pi} \right)^n \mathcal{H}^{'V(n)}_N \\[10pt]
& + \sum\limits_{n,m=1}^{\infty} \left( \dfrac{\as}{\pi} \right)^n 
\left( \dfrac{\alpha}{\pi} \right)^m \mathcal{H}'^{V(n,m)}_N  \, .
\end{array}
\end{equation}
The pure QED functions $g^{'(1)}$ and $g^{'(2)}$, as well as the first non-vanishing mixed \qcdqed~ 
\\
term, $g^{'(1,1)}$, are known in literature \cite{Autieri:2023xme,Cieri:2018sfk}. The functions $g^{'(2,1)}$, $g^{'(1,2)}$ and $g^{'(3)}$ are required to consistently achieve full mixed \qcdqed~NNLL accuracy. The computation of these three functions constitutes an original result of this paper.

Moving forward to the extension of the $q_T$-resummation formalism, the integral representation of $\mathcal{G}_N$ including \qcdqed~corrections is given by
\begin{equation}
    \mathcal{G}_N = -\int_{b_0^2/b^2}^{Q^2} \frac{dq^2}{q^2} \left[ A( \as(q^2),\alpha(q^2) ) \log \left( \frac{M^2}{q^2} \right) + \widetilde{B}_N ( \as(q^2), \alpha(q^2)) \right] ,
\end{equation}
where $A( \as(q^2), \alpha(q^2))$ and $\widetilde{B}_N( \as(q^2), \alpha(q^2))$ are double perturbative expansions in $\as$ and $\alpha$, defined according to
\begin{equation*}
    A( \as, \alpha) = \frac{\as}{\pi} A^{(1)} + \left( \frac{\as}{\pi} \right)^2 A^{(2)} + \sum_{n=3}^{\infty} \left( \frac{\as}{\pi} \right)^n A^{(n)} 
\end{equation*}
\begin{equation*}
    + \frac{\alpha}{\pi} A^{'(1)} + \left( \frac{\alpha}{\pi} \right)^2 A^{'(2)} + \sum_{n=3}^{\infty} \left( \frac{\alpha}{\pi} \right)^n A^{'(n)} 
\end{equation*}
\begin{equation}
    + \frac{\as}{\pi} \frac{\alpha}{\pi} A^{(1,1)} + \left( \frac{\as}{\pi} \right)^2 \frac{\alpha}{\pi} A^{(2,1)} + \frac{\as}{\pi} \left( \frac{\alpha}{\pi} \right)^2 A^{(1,2)} + \sum_{n,m=2}^{\infty}  \left( \frac{\as}{\pi} \right)^n \left( \frac{\alpha}{\pi} \right)^m A^{(n,m)} \, ,
\end{equation}
and
\begin{equation*}
    \widetilde{B}_N( \as, \alpha ) = \frac{\as}{\pi} \widetilde{B}_N^{(1)} + \left( \frac{\as}{\pi} \right)^2 \widetilde{B}_N^{(2)} + \sum_{n=3}^{\infty} \left( \frac{\as}{\pi} \right)^n \widetilde{B}_N^{(n)} 
\end{equation*}
\begin{equation*}
    + \frac{\alpha}{\pi} \widetilde{B}_N^{'(1)} + \left( \frac{\alpha}{\pi} \right)^2 \widetilde{B}_N^{'(2)} + \sum_{n=3}^{\infty} \left( \frac{\alpha}{\pi} \right)^n \widetilde{B}_N^{'(n)} 
\end{equation*}
\begin{equation}
    + \frac{\as}{\pi} \frac{\alpha}{\pi} \widetilde{B}_N^{(1,1)} + \left( \frac{\as}{\pi} \right)^2 \frac{\alpha}{\pi} \widetilde{B}_N^{(2,1)} + \frac{\as}{\pi} \left( \frac{\alpha}{\pi} \right)^2 \widetilde{B}_N^{(1,2)} + \sum_{n,m=2}^{\infty}  \left( \frac{\as}{\pi} \right)^n \left( \frac{\alpha}{\pi} \right)^m  \widetilde{B}_N^{(n,m)} \, .
\end{equation}
In the \qcdqed~formulation, we must consistently include the mixed \qcdqed~contributions to the running of the QCD and QED couplings simultaneously. As a consequence, the renormalization group equations (RGE) are written as \cite{Cieri:2018sfk, Billis:2019evv}: 
\begin{equation}
\label{rengroupQCD}
    \frac{d \ln{\as (\mu^2)}}{d \ln{\mu^2}} = \beta(\as,\alpha) = - \sum_{n=0}^{\infty} \beta_n \left( \frac{\as}{\pi} \right)^{n+1}  - \sum_{n,m+1=0}^{\infty} \beta_{n,m} \left( \frac{\as}{\pi} \right)^{n+1} \left( \frac{\alpha}{\pi} \right)^m \, ,
\end{equation}
\begin{equation}
\label{rengroupQED}
    \frac{d \ln{\alpha (\mu^2)}}{d \ln{\mu^2}} = \beta'(\as,\alpha) = - \sum_{n=0}^{\infty} \beta'_n \left( \frac{\alpha}{\pi} \right)^{n+1} - \sum_{m,n+1=0}^{\infty} \beta'_{m,n} \left( \frac{\alpha}{\pi} \right)^{m+1} \left( \frac{\as}{\pi} \right)^{n} \, ,
\end{equation}
with the inclusion of the mixed \qcdqed~$\beta$ functions: $\beta_{n,m}$ and $\beta'_{n,m}$. The known coefficients useful for our discussion are \cite{Cieri:2018sfk,Billis:2019evv}: 
\begin{equation}
    \beta_{1,1} = \frac{1}{16} ( C_F - 2 C_A) T_F N_q^{(2)} \, ,
\end{equation}
\begin{equation}
    \beta_0' = -\frac{N^{(2)}}{3}, \, \beta'_1 = -\frac{N^{(4)}}{4}, \, \beta'_{1,0} = - \frac{C_F C_A N_q^{(2)}}{4},
     \, \beta'_{1,1} = \frac{-C_F N_c}{16}N_q^{(2)} \, ,
\end{equation}
\begin{equation}
    \beta_{0,2} = \frac{1}{64} \left( \frac{44}{9} T_F N_q^{(2)} N^{(2)} + 2 \, T_F N_q^{(4)} \right) \, ,
\end{equation}
\begin{equation}
        \beta'_{2,0} = -\frac{1}{64} \left( \left(\frac{133}{18} C_A - C_F \right) 2 C_F N_c N_q^{(2)} - \frac{44}{9} C_F T_F N_f N_c N_{q}^{(2)}  \right) \, ,
\end{equation}
\begin{equation}
\beta'_2 = -\frac{1}{64} \left( \frac{44}{9} N^{(4) }N^{(2)}+2 N^{(6)} \right),    
\end{equation}
where we introduced the short-hand definitions
\begin{equation}
    N^{(n)} = N_c \sum_{i=1}^{n_f} e_{q_i}^n + \sum_{l=1}^{n_l} e_l^n,
\end{equation}
\begin{equation}
    N_q^{(n)} = \sum_{i=1}^{n_f} e_{q_i}^n \, ,
\end{equation}
with $N_c = 3$ the number of colors, $n_f$ ($n_l$) the number of quark (lepton) flavours and $e_q$ ($e_l$) the quark (lepton) electric charges.

At the lowest order in perturbation theory, RGE equations are decoupled. Then, the solution is given by 
\begin{equation}
  \as^{\rm LO}(Q^2) =   \frac{\as(\mu_R^2)}{y} \, ,
  \label{eq:aSLOQCD}
\end{equation}
with 
\beq
\lambda_\mu = \log ( Q^2/\mu^2)
\eeq
and
\begin{equation}
    y = 1 - \lambda_{\mu} \beta_0 \as ( \mu_R^2) \, ,
\end{equation}
where we explicitly introduced the renormalization scale $\mu_R$. Starting from the next perturbative order, the equations are coupled and the evolution of QCD coupling presents QED-induced corrections. If we use the expansion
\beq
\as \equiv \sum_{i,j=0} \delta \as^{(i,j)} \, ,
\eeq
with $\as^{\rm LO} \equiv \as^{(0,0)}$, we can write the NLO QCD corrections as
\begin{equation}
    \delta \as^{(1,0)}(Q^2) = - \as^2(\muRsq) \frac{\beta_1 \beta'_0 \ln{y} + \beta_{0,1} \beta_0 \ln{y'} + \beta_0^2 \beta_0' \ln{\frac{Q^2}{\mu_R^2}}}{\beta_0 \beta'_0 y^2} \, ,
    \label{eq:aSNLOQCD}
\end{equation}
where QED effects manifest through the mixing $\beta$-coefficients and
\begin{equation}
    y' = 1 - \lambda_{\mu} \beta'_0 \alpha ( \mu_R^2) \, .
\end{equation}
The mixed NLO QCD+LO QED correction reads
\begin{align}
    \delta \alpha_{S}^{(1,0)} (Q^2) = \alpha (\muRsq) \as^2 (\muRsq) \frac{1}{\beta_0^2 \beta_0^{'2} y^3 y'} \biggl( ( \beta_{1,1} \beta_0 - \beta_1 \beta_{0,1}) \beta_0^{'2} y \, y' \ln{y} \nonumber
    \\
    + ( \beta_{0,1} \beta_0^2 \beta'_1 y + \beta_0 \beta'_0 (\beta_{0,1} \beta'_{1,0} - \beta_{1,1} \beta'_0 ) y \, y' + \beta_1 \beta_{0,1} \beta_0^{'2} y^{'2} ) \ln{y'}
    \nonumber 
    \\
    + \beta_0^2 y \left( (\beta_{0,1} \beta'_1 - \beta_{0,2} \beta_0')(y'-1) + \beta_{0,1} \beta_0^{'2} \ln{\frac{Q^2}{\mu_R^2}} \right) \biggl) \, ,
    \label{eq:aSNLOQCDLOQED}
\end{align}
whilst the LO QCD+NLO QED contribution is given by
\begin{align}
    \delta \alpha_{S}^{(0,1)} (Q^2) = \as(\muRsq) \alpha^2 (\muRsq)
    \frac{1}{\beta_0^3 y^3 y'} \biggl( y \, ( \beta_{0,1} \beta_0 \beta'_{1,0} y + \beta_1 \beta_{0,1} \beta'_0 y' - \beta_{1,1} \beta_0 \beta'_0 y' ) \ln{y} 
    \nonumber
    \\
    - y' (\beta_{0,1} \beta_0 \beta'_{1,0} y - \beta_{1,1} \beta_0 \beta'_0 y + \beta_1 \beta_{0,1} \beta'_0 y') \ln{y'} \biggl) \, .
    \label{eq:aSLOQCDNLOQED}
\end{align}
The NNLO QCD contribution to the evolution of the QCD coupling is 
\begin{align}
    \delta \as^{(2,0)} (Q^2) = \as^3 (\muRsq) \frac{1}{\beta_0^2 \beta_0^{'2} y^3} \biggl( (\beta_1^2 - \beta_2 \beta_0) \beta_0^{'2} (y-1) + \beta_1^2 \beta_0^{'2} \ln{y}^2 + \beta_0 \biggl( \beta_{0,1} \ln{y'} \nonumber 
    \\
    ( - \beta_{1} \beta'_0 y' + \beta_{0,1} \beta_0 \ln{y'})
    \nonumber
    + \beta_0 \beta'_0 ( -\beta_1 \beta_0' + 2 \beta_{0,1} \beta_0 \ln{y'}) \ln{\frac{Q^2}{\mu_R^2}} + \beta_0^3 \beta_0^{'2}\ln{\frac{Q^2}{\mu_R^2}}^2 \biggl)
    \nonumber
    \\
    +\beta_1 \beta_0' \log y \left( -\beta_1 \beta_0' + 2 \beta_0 \left( \beta_{0,1} \ln{y'} + \beta_0 \beta_0' \ln{\frac{Q^2}{\mu_R^2}} \right) \right) \biggl) \, ,
    \label{eq:aSNNLOQCD}
\end{align}
and, as expected, also includes mixed \qcdqed~corrections. The expressions given in Eqs. (\ref{eq:aSNLOQCD})–(\ref{eq:aSNNLOQCD}) have been recomputed independently by us, integrating the RGE. The procedure is analogous to the one described in Ref. \cite{Billis:2019evv}, and we find agreement with Eqs. (2.15)–(2.16) therein. Here, we used the short-hand notation
\begin{equation}
    \beta_{n,0} \equiv \beta_n, \, \, \beta'_{0,m} \equiv \beta'_m \, .
\end{equation}
The evolution of the QED coupling at $\mathcal{O} (\alpha)$, $\mathcal{O} (\alpha^2)$, $\mathcal{O} (\as\alpha^2)$, $\mathcal{O} (\as^2\alpha)$ and $\mathcal{O} (\alpha^3)$ can be obtained through the application of the replacements 
\begin{equation}
    \beta_{n,m} \leftrightarrow \beta'_{m,n}, \, \, \as (\mu_R^2) \leftrightarrow  \alpha (\mu_R^2)  \, ,
\end{equation} 
to Eqs. (\ref{eq:aSLOQCD}), (\ref{eq:aSNLOQCD}), (\ref{eq:aSNLOQCDLOQED}), (\ref{eq:aSLOQCDNLOQED}) and (\ref{eq:aSNNLOQCD}), respectively.

Returning to the expansion of ${\cal G}_N$ in Eq. (\ref{eq:gmixedcoefs}), the explicit forms of the functions $g^{'(2,1)}$, $g^{'(1,2)}$ and $g^{'(3)}$ are inferred from the integral representation of the universal form factors, combining the coefficients $\widetilde{B}_q^{(1,1)}$, $A_q^{(2,1)}$, $A_q^{(1,2)}$, $\widetilde{B}_q^{'2}$ and $A_q^{'3}$ together with the evolution of the couplings including up to \qcdqed~NNLO corrections. $\widetilde{B}_q^{(1,1)}$ and  $\widetilde{B}_q^{'(2)}$ are given by:
\begin{equation}
    \widetilde{B}_{q,N}^{(1,1)} = B_q^{(1,1)} -  \beta_0C^{'(1)}_{qq,N} - \beta_0'C^{(1)}_{qq,N} + 2 \gamma_{qq,N}^{(1,1)}  
\end{equation}
and: 
\begin{equation}
    \widetilde{B}_{q,N}^{'(2)} = B_q^{'(2)} -2 \beta_0'C^{'(1)}_{qq,N} + 2 \gamma_{qq,N}^{'(2)}  \, ,
\end{equation}
where $\gamma_{qq,N}^{(1,1)}$ and $\gamma_{qq,N}^{'2}$ are respectively the mixed $\alpha_S \alpha$ and the second order QED Mellin moments of the Alterelli Parisi splitting kernels \cite{deFlorian:2016gvk, deFlorian:2015ujt}. $B_q^{(1,1)}$ and  $B_q^{'(2)}$, instead, are obtained from $B_q^{(2)}$ by applying the Abelianization rules from Table 1 of Ref.~\cite{deFlorian:2018wcj}, which leads to
\begin{equation}
    B_q^{'(2)} = N^{(2)} e_q^2 \left(\frac{17}{12} - \frac{\pi^2}{9}\right) + e_q^4 \left(\frac{\pi^2}{4} - \frac{3}{16} - 3 \zeta_3\right) \, ,
\end{equation} 
whilst $B_q^{(1,1)} $ is given in Eq.~(23) of Ref.~\cite{Cieri:2020ikq}. As discussed in Ref. \cite{Catani:2013tia}, $B_q^{(n)}$ is scheme dependent for $n\geq2$ and we are choosing the DY-scheme.

Since we consider the next order in mixed \qcdqed, the expressions for $A_q^{(2,1)}$, $A_q^{(1,2)}$ and $A_q^{'(3)}$ are necessary. These coefficients are directly related to the cusp anomalous dimensions, which were extracted from the form factor explicitly evaluated up to three loops in Ref.~\cite{AH:2019pyp}. We independently obtained the cusp anomalous dimensions through the Abelianization of the corresponding QCD diagrams shown in Table 2 of Ref. \cite{AH:2019pyp}, finding total agreement with the expressions given by the authors. Due to the fact that $A_q^{(3)}$ is observable-dependent, we use the definition given in Ref. \cite{Becher:2010tm} for the $A_q$ coefficient in pure QCD, i.e.
\beq
A^{(3,0)} = \frac{1}{64}\left[\Gamma_F^{(2)} + 2 \beta_0 \, d_2^q \right]\, ,
\label{eq:DefineA3}
\eeq
with
\beqn
\nonumber \Gamma_F^{(2)} &=& C_A^2 C_F \left(\frac{490}{3}-\frac{1072}{9}\zeta_2+\frac{44}{45}\pi^4+\frac{88}{3}\zeta_3 \right) +C_F C_A T_F n_F \left(-\frac{1672}{27}+\frac{320}{9}\zeta_2-\frac{224}{3}\zeta_3\right) \,
\\ &+& C_F^2 T_F n_F \left(-\frac{220}{3}+64\zeta_3\right) +C_F T_F^2 n_F^2 \left(-\frac{64}{27}\right) \, ,
\label{eq:GammaF2}
\eeqn
and
\beqn
d_2^q &=& C_A C_F \left(\frac{808}{27} - 28 \zeta_3\right) +C_F T_F n_F \left(-\frac{224}{27} \right) \, ,
\eeqn
where we adjusted the normalization in Eq. (\ref{eq:DefineA3}) in order to match the one originally present in \texttt{DYTurbo}. Putting the ingredients altogether, we finally obtain:
\begin{equation}
    A^{(2,1)} = C_F T_F N_q^{(2)} \left( \zeta_3 - \frac{55}{48} \right) \, ,
\end{equation}
\begin{equation}
    A^{(1,2)} = C_F e_q^2 N_C N_q^{(2)} \left( \zeta_3 - \frac{55}{48} \right) \, ,
\end{equation}
and
\begin{equation}
    A^{(0,3)} = A'^{(3)} = e_q^2 N^{(4)} \left( \zeta_3 - \frac{55}{48} \right) + e_q^2(N^{(2)})^2 \frac{25}{162}   \, .
\end{equation}
We appreciate that $A^{(2,1)}$ and $A^{(1,2)}$ are equal to the cusp anomalous dimensions, but the second term in $A^{(0,3)}$ exhibits a different behaviour due to observable-dependence induced by Eq. (\ref{eq:DefineA3}). Furthermore, we note that the cusp and collinear anomalous dimensions at third order in QCD, from which we extracted the mixed counterpart, were also computed independently in Ref.~\cite{Stadlbauer:2024jij}.

With all these coefficients, we can describe up to \qcdqed~NNLL corrections that resum terms proportional to $\as^n \alpha^m L^{n+m}$, which are embodied in both functions $g_N^{'(2,1)}$ and $g_N^{'(1,2)}$. These functions can be symbolically written as
\begin{equation}
    \as \, g_N^{'(2,1)} =  f (\lambda, \lambda') \, ,
\end{equation}
\begin{equation}
     \alpha \, g_N^{'(1,2)} =  f' (\lambda, \lambda') \, ,
\end{equation}
with
\begin{equation}
    \lambda = \frac{1}{\pi} \beta_0 \as L \, ,
\end{equation}
\begin{equation}
    \lambda' = \frac{1}{\pi} \beta_0' \alpha L \, ,
\end{equation}
although their explicit analytic form is quite complex, so we report them as ancillary files \cite{ZENODO}. From the ratio of $\lambda$ to $\lambda'$, it is possible to show that we have the freedom to move terms from $g_N^{'(2,1)}$ to $g_N^{'(1,2)}$: only their sum is consistent in fully realizing the mixed \qcdqed~NNLL correction.

The last ingredient to describe ${\cal G}_N$ is $g_N^{'(3)}$. This function is straightforwardly obtained by a naive Abelianization of its QCD counter-part (i.e. $g_N^{(3)}$ reported in Ref. \cite{Bozzi:2005wk}) and it is given by
\begin{align}
    & g_N'^{(3)} \left( \alpha L; \frac{M^2}{\mu_R^2}, \frac{M^2}{Q^2} \right) 
     = - \frac{A'^{(3)}}{2 \beta_0^{'2}} \frac{\lambda'^{2}}{ (1-\lambda'^{2})} 
    - \frac{\bar{B}_N'^{(2)}}{\beta_0'}\frac{\lambda'}{1-\lambda'} \nonumber \\
    & \quad + \frac{A'^{(2)}\beta_1'}{\beta_0'^{3}} 
    \left( \frac{\lambda' (3 \lambda' -2)}{ 2 (1-\lambda')^2} 
    - \frac{ (1-2 \lambda') \log (1-\lambda')}{ (1-\lambda'^2)}\right) \nonumber \\
    & \quad + \frac{\bar{B}_N'^{(1) } \beta'_1}{\beta_0'^2} \left( \frac{\lambda'}{1-\lambda'} + \frac{\log(1-\lambda')}{1-\lambda'} \right) \nonumber \\
    & \quad - \frac{A'^{(1)}}{2} \frac{\lambda'^2}{ (1-\lambda')^2} \log^2 \frac{Q^2}{\mu_R^2}  \nonumber \\
    & \quad + \log \frac{Q^2}{\mu_R^2} \left( \bar{B}_N'^{(1)} \frac{\lambda'}{ 1-\lambda'} + \frac{A'^{(2)}}{\beta_0'} \frac{\lambda'^2}{ (1-\lambda')^2 } 
    +A'^{(1)} \frac{\beta_1'}{\beta_0'^2} \left( \frac{\lambda'}{1-\lambda'} + \frac{1-2 \lambda'}{(1-\lambda')^2} \log (1-\lambda')  \right)\right) \nonumber \\
    & \quad + A'^{ (1)} \biggl( \frac{\beta_1'^2}{2 \beta_0'^4} \frac{1- 2 \lambda'}{ (1-\lambda')^2} \log^2 (1-\lambda') 
    + \log (1-\lambda') \left[ \frac{\beta_0' \beta_2' - \beta_1'^2}{\beta_0'^4} + \frac{\beta_1'^2}{\beta_0'^4 (1-\lambda')} \right] \nonumber \\
    & \quad + \frac{\lambda'}{2 \beta_0'^4 (1-\lambda')^2} ( \beta_0' \beta_2' (2- 3 \lambda') + \beta_1'^2 \lambda' ) \biggl) \, ,
\end{align}
where we introduced
\beq
\bar{B}_N^{(n)} = \widetilde{B}_N^{(n)} + A^{(n)} \, \log\left(\frac{M^2}{Q^2}\right) \, .
\eeq

Finally, we move our discussion to the hard collinear coefficient function $\mathcal{H}'^{F}_N$ including mixed $\mathcal{O} (\as \alpha)$ and $\mathcal{O} (\alpha^2)$ corrections.
The function $H^{Z \,(1,1)}$ and the collinear coefficients $C_{ab}$ are extracted from Eqs.~(26)–(30) of Ref.~\cite{Cieri:2020ikq}. They were obtained from the QCD ones \cite{Catani:2012qa} by applying the Abelianization algorithm from Ref. \cite{deFlorian:2018wcj}. Analogously, $\mathcal{H}'^{Z(2)}_N$ was also computed using this Abelianization procedure.



\section{Numerical results}
\label{sec:ResultsZ}
In this Section, we present a phenomenological study of $Z$ and $\gamma^*$ production at LHC and Tevatron. For this purpose, we rely on the numerical code \texttt{DYTURBO}~\cite{Camarda:2019zyx,Camarda:2021ict,Camarda:2021jsw,Camarda:2023dqn}, where we have encoded the full mixed \qcdqed~resummed corrections that we have calculated in this paper. To be more precise in this context, \emph{NLL mixed corrections} account for the pure QED resummation at NLL plus the resummation of the mixed logarithms of the form
\beq
(\as^n\alpha^m/q_T^2) \log^{n+m} \left(\frac{M^2}{q_T^2}\right) \, ,
\eeq
while \emph{NNLL mixed corrections} include pure QED resummation at NNLL and the resummation of the mixed logarithms of the form  
\beq
(\as^n\alpha^m/q_T^2) \log^{n+m+1} \left(\frac{M^2}{q_T^2}\right) \, ,
\eeq
with $M$ the invariant mass of the final state.

As our default setup, we use the input parameters $\alpha (m_Z^2) = 1/128.89$, $\sin^2(\theta_W) = 0.23129$ and $m_Z= 91.6660 \, {\rm GeV}$, together with $G_F = 1.16496 \times 10^{-5} \, \text{GeV}^{-2}$ (extracted from the tree-level EW constraints) and the PDF set \texttt{NNPDF40.QED} \cite{NNPDF:2024djq} which includes $\mathcal{O} (\alpha)$, $\mathcal{O} (\as \alpha)$ and $\mathcal{O} (\alpha^2)$ corrections to the QED evolution. Regarding the strong coupling, we used 3 and 4 loop corrections, respectively, for NNLO and $\text{N}^3\text{LO}$ QCD predictions, with $\as (m_Z^2) = 0.118$ in the $\overline{\text{MS}}$ renormalization scheme. Finally, we consider $n_f = 5$ (number of light quarks flavours) and $n_l = 3$ (number of charged leptons) in the massless approximation.

In order to assess the uncertainty due to missing higher-order terms in QCD and QED, the scale variation method is used. We set the central scale to $Q = \mu_F =  \mu_R = M$ and we vary the scales according to the following constraint: 
\begin{equation}
    M/2 \leq \mu_F, \mu_R, 2Q \leq 2M \, .
\end{equation}

\begin{figure}[h!]
    \centering
    \includegraphics[width=0.49
\linewidth]{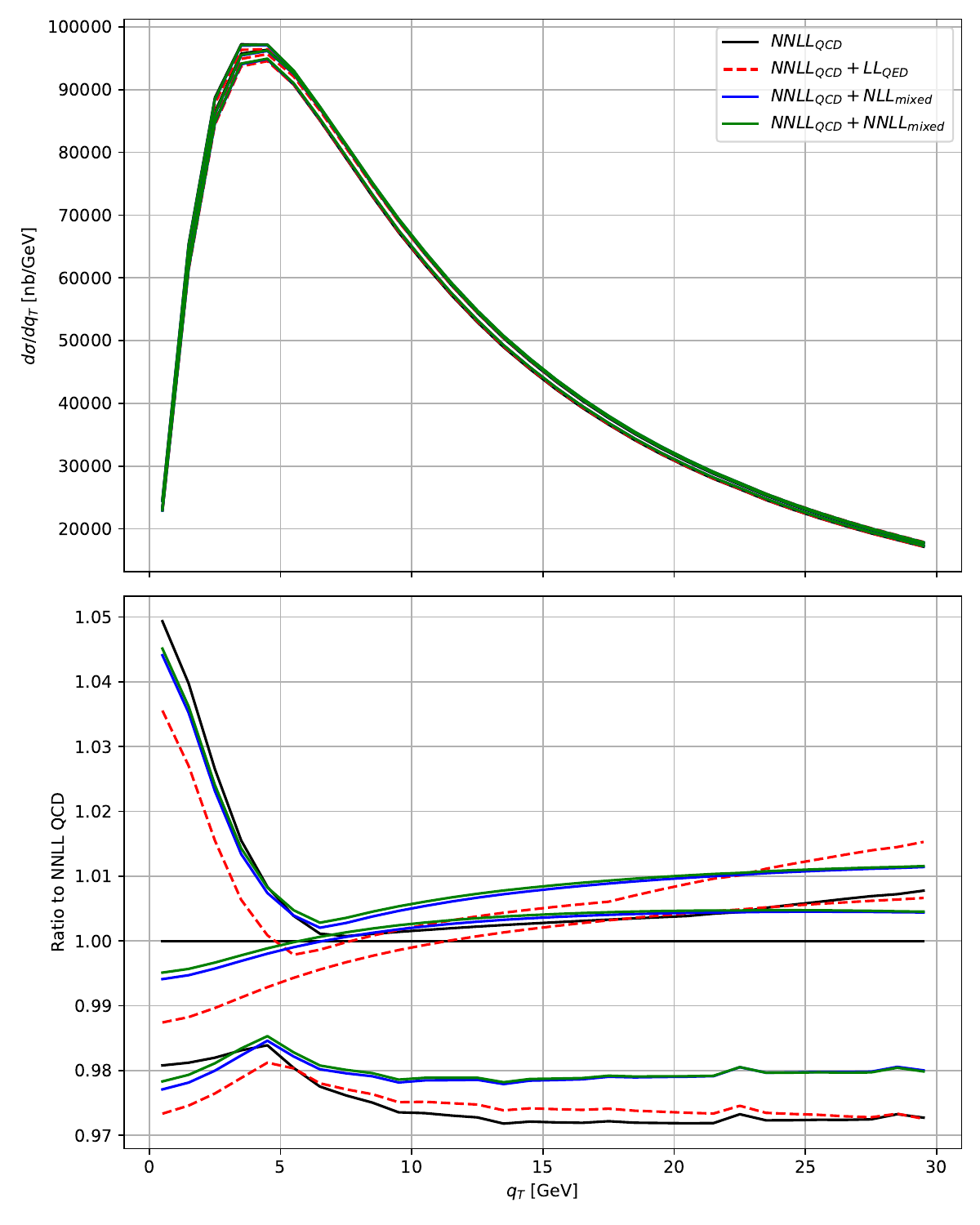}
    \includegraphics[width=0.49\linewidth]{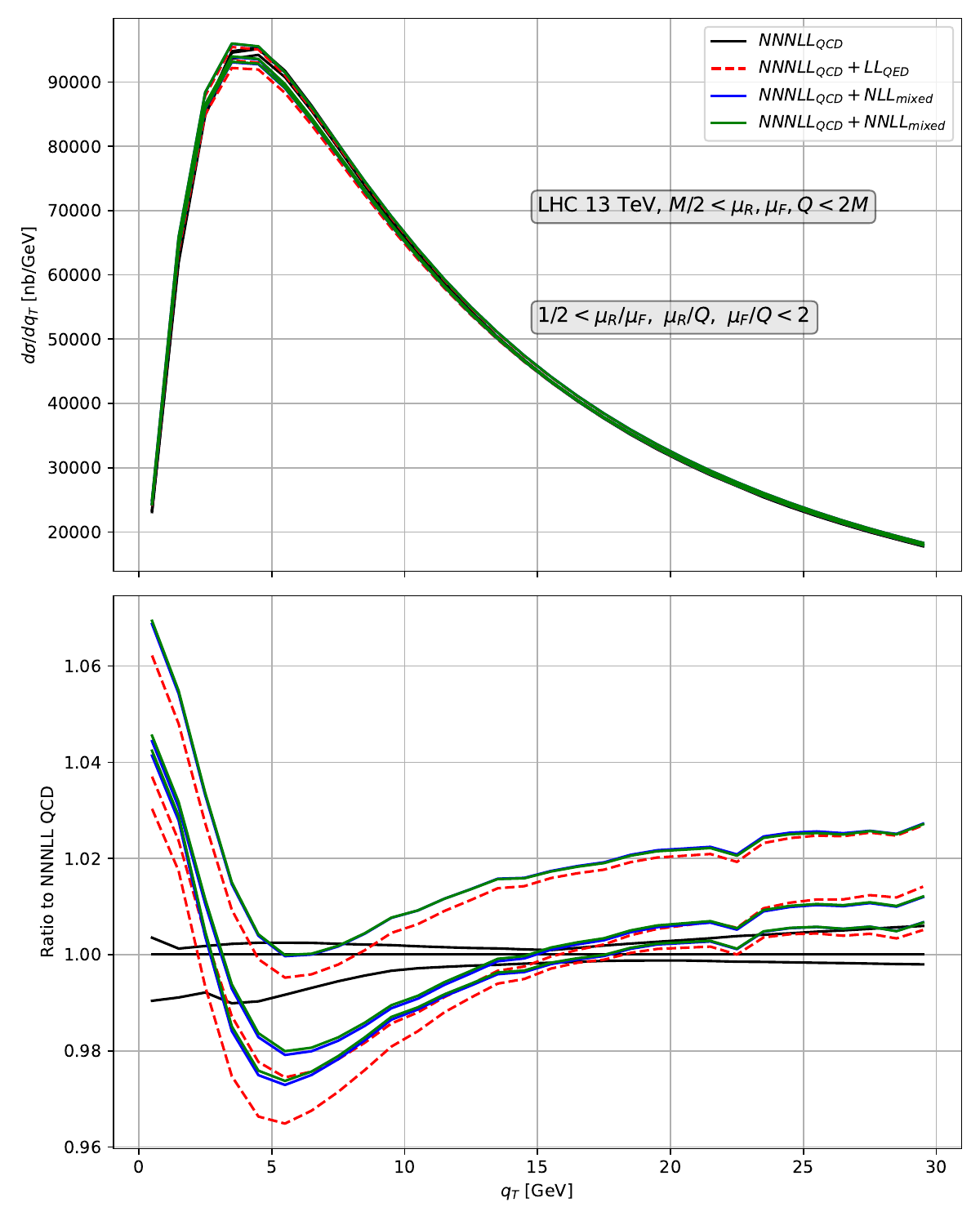}
    \caption{$Z$/$\gamma^*$ production at LHC, $\sqrt{s} = 13 \, \text{TeV}$ at NNLL (left plot) and N$^3$LL (right plot) in QCD with the inclusion of mixed \qcdqed~effects up to NNLL. In the upper plot the differential distributions are showed, while in the lower one we show the ratios with respect to the QCD prediction at the central scale (black line).}
    \label{plot1}
\end{figure}

In Fig. \ref{plot1}, we consider the $q_T$ distribution for $Z$ and $\gamma^*$ production at LHC at $\sqrt{s}=13$ TeV, including up to NNLL mixed \qcdqed~effects on top of NNLL (left plot) and N$^3$LL (right plot) QCD contributions (black line). In both cases, the peak of the distribution is located at $q_T \approx 4$ GeV. For the central scale predictions, compared to pure QCD at NNLL, we observe that mixed corrections at NLL and NNLL make the spectrum harder for $q_T > 5$ GeV, while LL corrections make the spectrum harder for $q_T > 10 $ GeV. Respectively to pure QCD at N$^3$LL, instead, we see that mixed and pure QED corrections are positive for $q_T < 3 $ GeV and $q_T > 15 $ GeV, whilst negative in the middle. Also for the central scale, we observe that LL QED (red line) effects are  ${\cal O}(0.5\%)$ for $q_T > 10$ GeV and ${\cal O}(-1\%)$ for $q_T < 10$ GeV when compared to NNLL QCD predictions, while they range from $\mathcal{O } (-2 \%)$ to $\mathcal{O} (4 \%)$ compared to N$^3$LL QCD prediction. The mixed NLL (blue line) and NNLL (green line) corrections are very similar, differing at the permille level. 

Regarding error bands in Fig. \ref{plot1}, compared to NNLL QCD (left), we appreciate that the scale variation is ${\cal O}(5 \%)$ for all cases, indicating that the uncertainty is mainly driven by pure QCD contributions. Furthermore, the error band compared to N$^3$LL QCD (right) is reduced with respect the previous, reaching ${\cal O}(2 \%)$, whilst pure QCD effects lead to ${\cal O}(1 \%)$. This supports the fact that QED and mixed effects do not affect the convergence of the QCD series.

Besides this, it is worth highlighting that mixed NLL and NNLL are shifted to the right (i.e. the hardest part of the spectrum) w.r.t. LL QED corrections. In fact, the ratio to the QCD predictions (both at NNLL) becomes 1 at $q_T \sim 10$ GeV for LL QED and at $q_T=5$ GeV for mixed NLL and NNLL QED. Similarly, even if the error bands have roughly the same width, LL QED is partly shifted down with respect to mixed NLL and NNLL.

\begin{figure}[h!]
    \centering
    \includegraphics[width=0.49\linewidth]{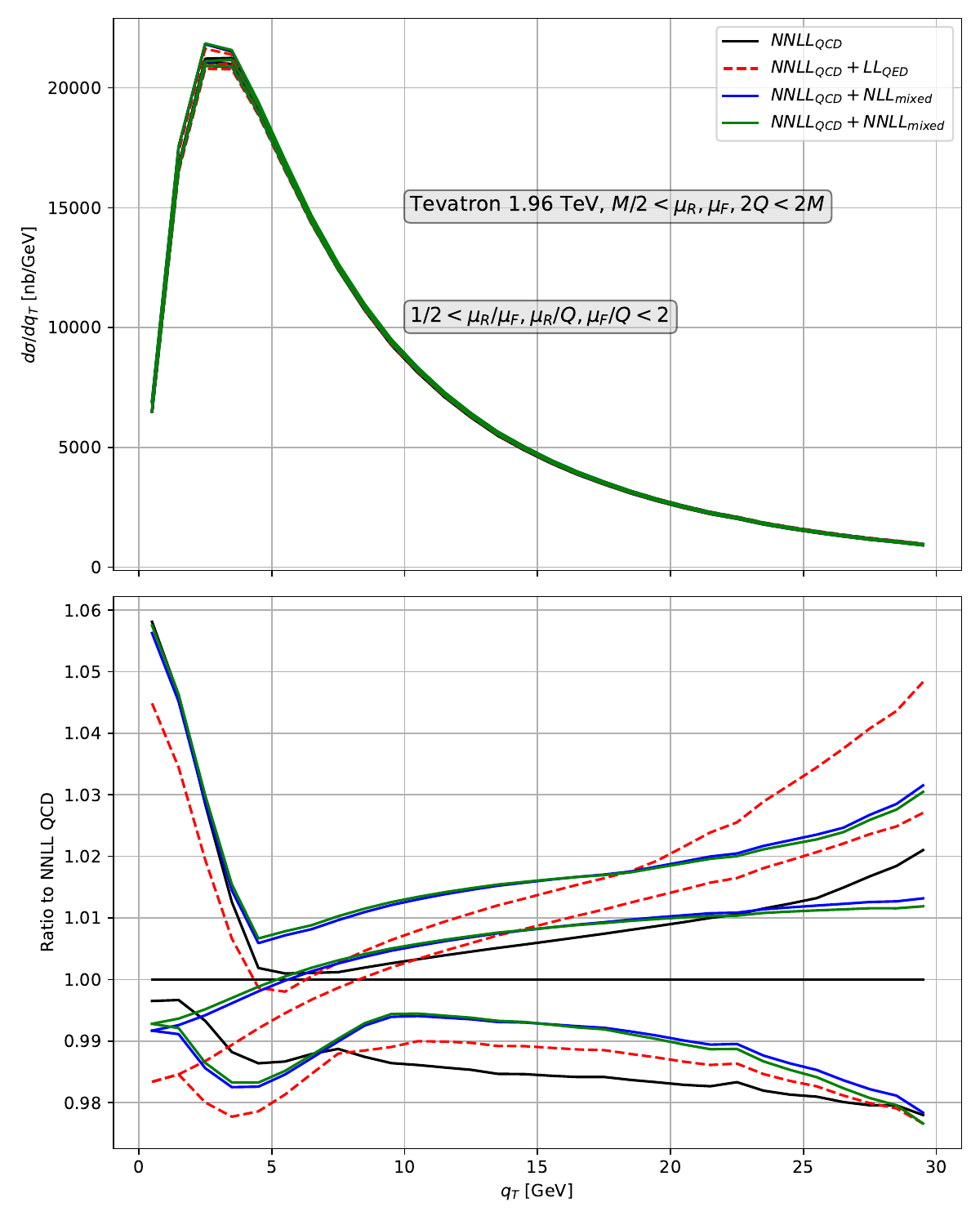}
        \includegraphics[width=0.49\linewidth]{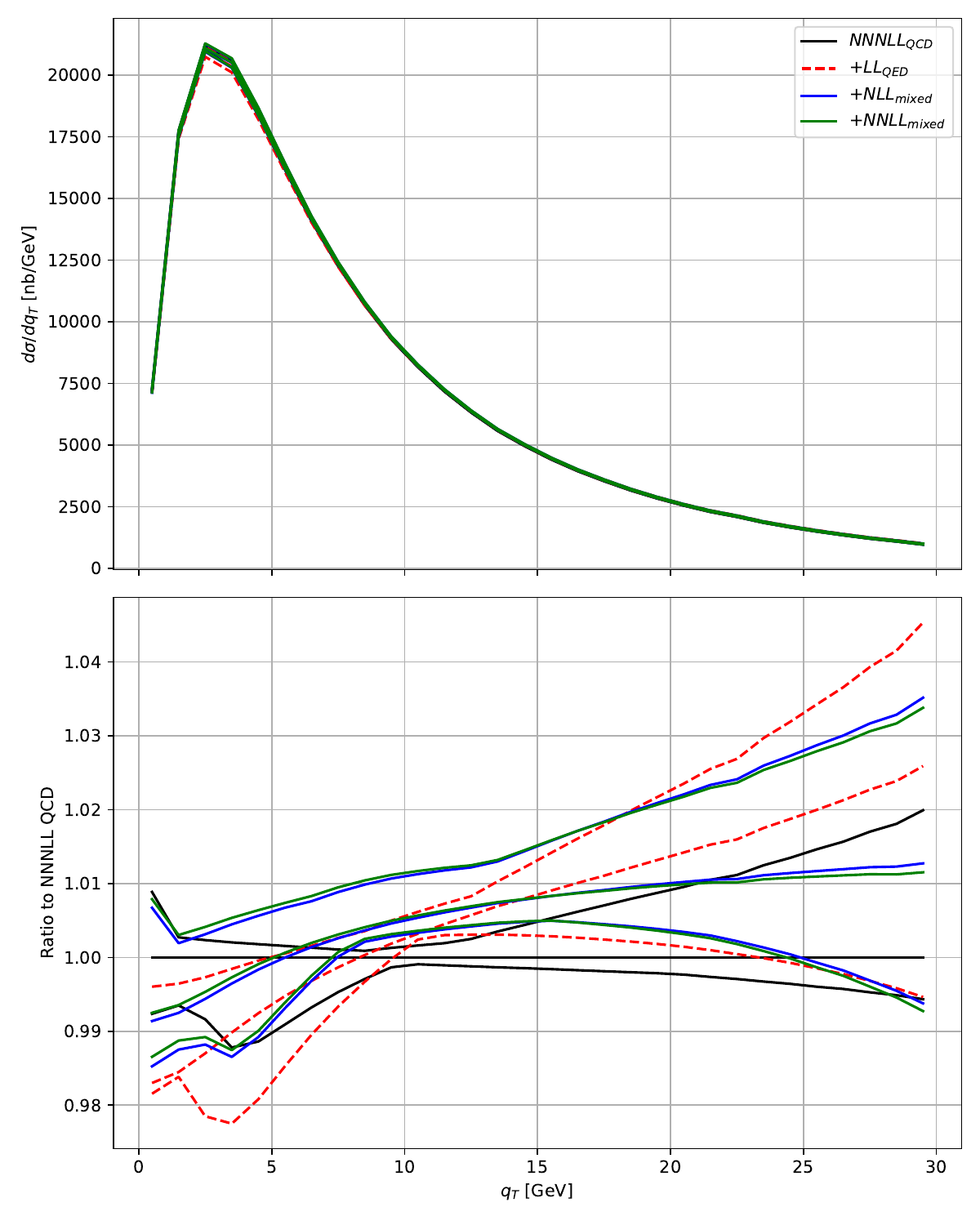}
    \caption{$Z$/$\gamma^*$ production at Tevatron, $\sqrt{s} = 1.96 \, \text{TeV}$ at NNLL (left plot) and N$^3$LL (right plot) in QCD with the inclusion of mixed \qcdqed~effects up to NNLL. In the upper plot the differential distributions are showed, while in the lower one we show the ratios with respect to the QCD prediction at the central scale (black line).}
    \label{plot3}
\end{figure}

In Fig. \ref{plot3}, we show the $q_T$ spectrum for $Z$ and $\gamma^*$ production at the Tevatron at $\sqrt{s} = 1.96 \, \text{TeV}$, including NNLL (left) and N$^3$LL (right) QCD resummation and up to mixed NNLL effects. Due to the suppression of gluon-induced reactions, the effect of QED and mixed corrections is enhanced, reaching up to ${\cal O}(2\%)$ for LL QED and ${\cal O}(1\%)$ for mixed NLL and NNLL corrections. As in the LHC case, the insertion of QED and mixed terms makes the spectrum harder, and the behavior is similar when considering the ratios with respect to the NNLL (left) and N$^3$LL QCD (right) contributions (black line). The scale variation band is ${\cal O}(3\%)$, and a slight overlap of the bands is observed for $q_T > 15\,\text{GeV}$, indicating that QED and mixed uncertainties start to be comparable with QCD ones, showing a good convergence. In fact, the pure QCD error band is ${\cal O}(3\%)$ at NNLL (left) and ${\cal O}(1\%)$ at N$^3$LL (right).

\begin{figure}[htbp]
    \centering
    \includegraphics[width=0.48\linewidth]{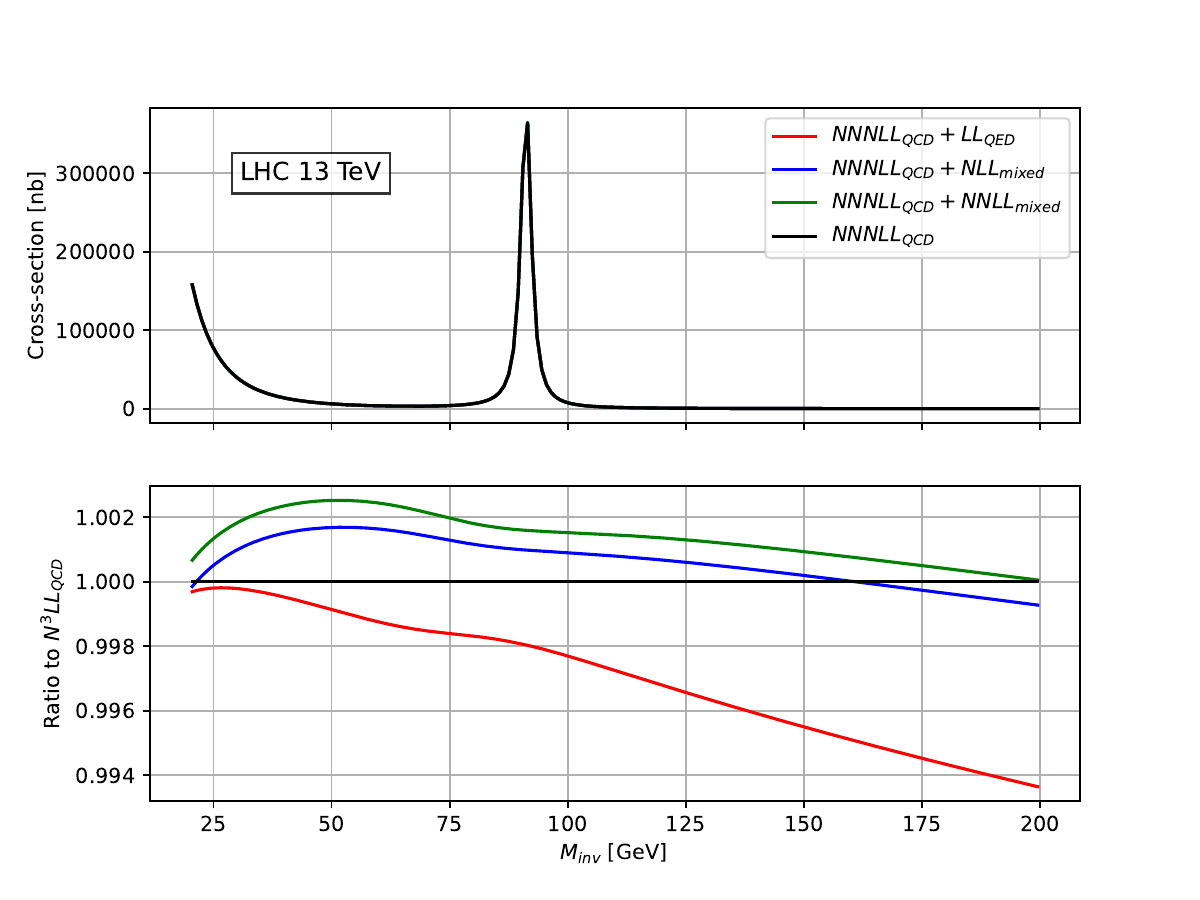}
    \includegraphics[width=0.48\linewidth]{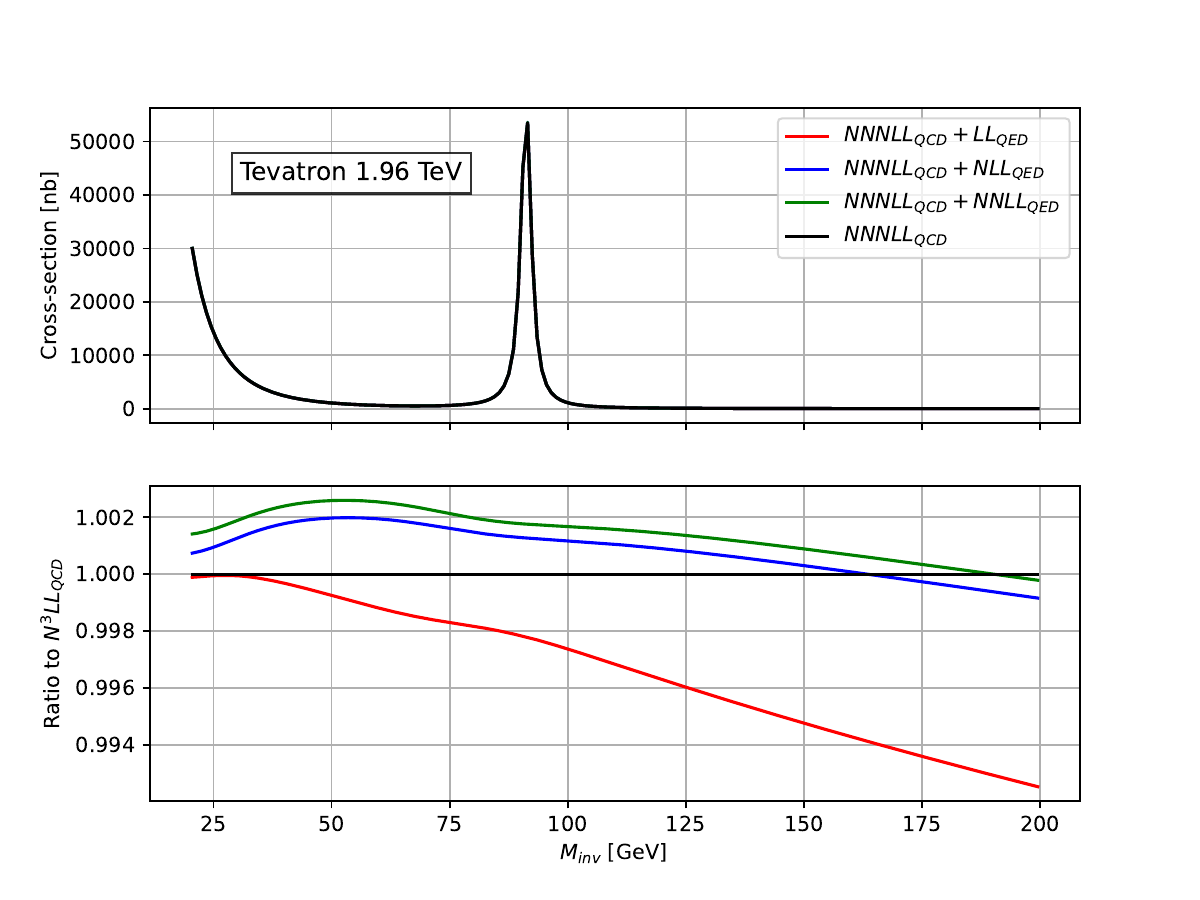}
    \caption{$Z$/$\gamma^*$ production at the LHC with $\sqrt{s} = 13 \, \text{TeV}$ (left) and Tevatron with $\sqrt{s} = 1.96 \, \text{TeV}$ (right) at $\text{N}^3\text{LL}$ in QCD with the inclusion of mixed effects up to NNLL. In the upper plot the differential distributions in the invariant mass are showed, while in the lower one the ratios with respect to the QCD prediction (black line).} 
    \label{plot5}
\end{figure}

After the analysis of the $q_T$ spectrum, we consider the invariant mass plot at the LHC (left) and Tevatron (right), as shown in Figure \ref{plot5}. For the LHC (left plot), the effect of QED corrections at LL around the peak is pronounced, lowering the distribution by approximately $0.2\%$. Mixed NLL and NNLL corrections drive the distribution in the opposite direction, increasing the distribution at the peak by roughly $\sim 0.1\%$ and $\sim 0.2\%$, respectively. This qualitative behavior is rather similar for Tevatron (right plot), with an overall enhanced QED impact due to the suppression of gluon PDF.

\begin{figure}[htbp]
    \centering
    \includegraphics[width=0.48\linewidth]{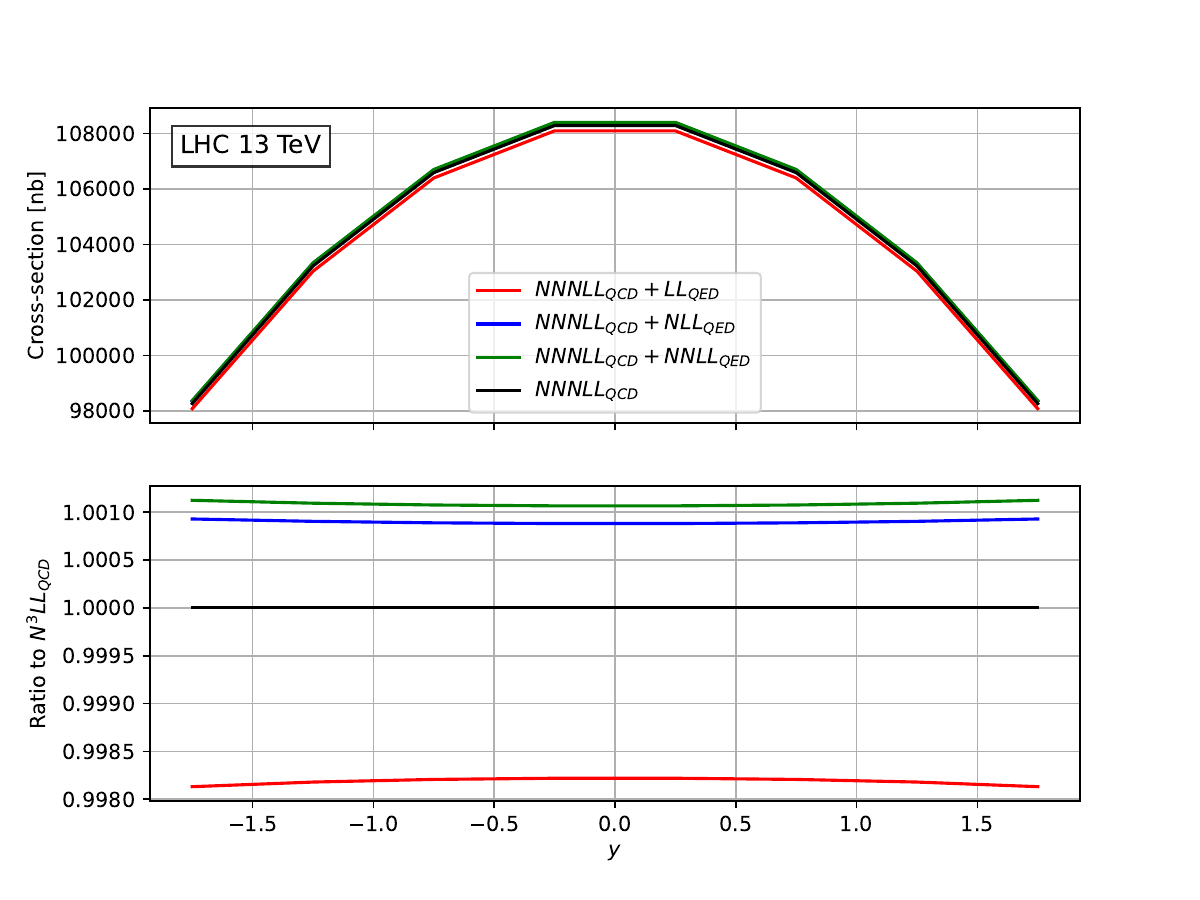}
    \includegraphics[width=0.48\linewidth]{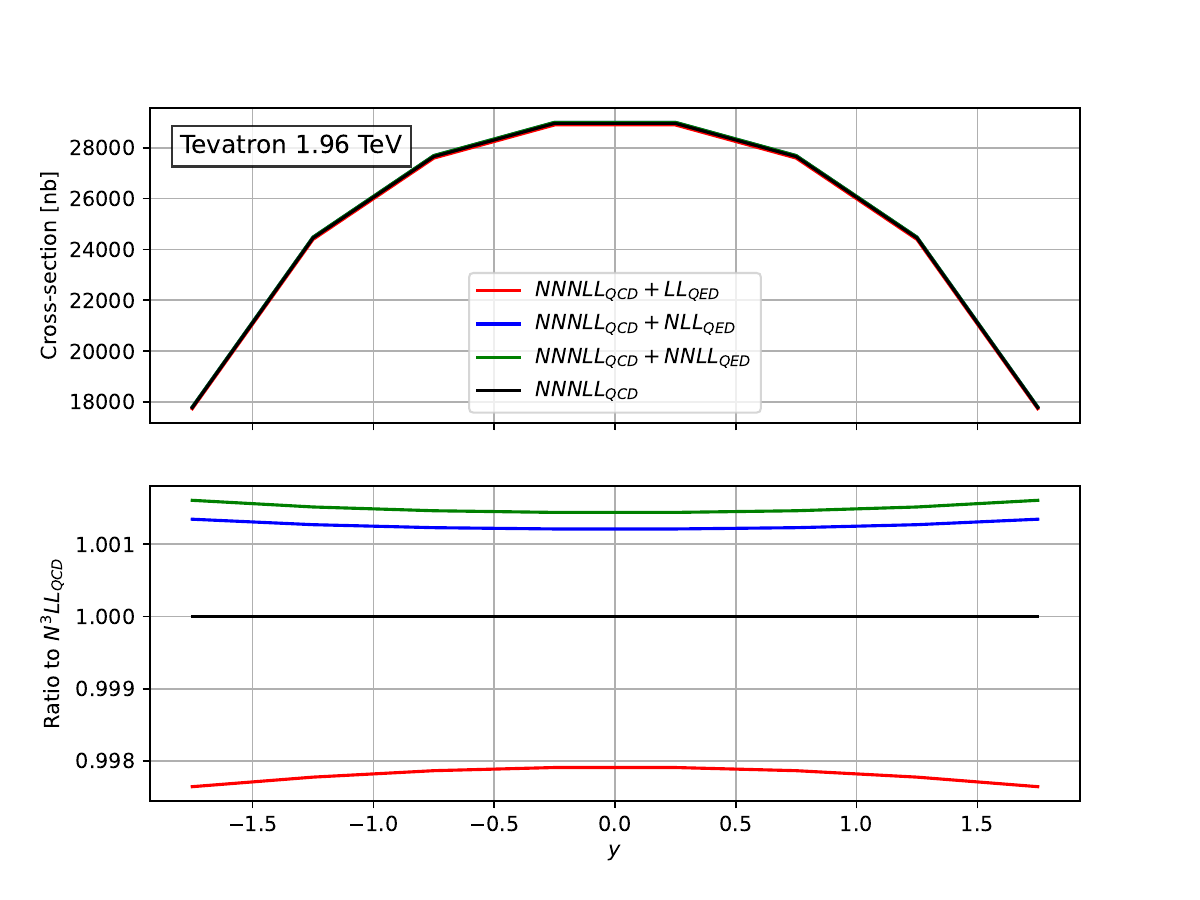}
    \caption{$Z$/$\gamma^*$ production at the LHC with $\sqrt{s} = 13 \, \text{TeV}$ (left) and Tevatron with $\sqrt{s} = 1.96 \, \text{TeV}$ (right) at $\text{N}^3\text{LL}$ in QCD with the inclusion of mixed effects up to NNLL. In the upper plot the differential distributions in the rapidity $y$ are showed, while in the lower one the ratios with respect to the QCD prediction (black line).} 
    \label{plot6}
\end{figure}

As a final phenomenological study, we present the rapidity distribution \( y \) at the LHC (left) and the Tevatron (right) in Fig.~\ref{plot6}.  
At the LHC, the effect of QED radiation at LL is relatively constant, resulting in a downward shift of the spectrum by approximately \( \sim 0.15\% \).  
In contrast, the mixed NLL and NNLL corrections lead to an increase in the spectrum by about \( \sim 0.1\% \), with a slightly more pronounced effect for the NNLL correction. The numerical predictions at the Tevatron show a similar qualitative behavior, although less uniform, with enhanced QED and mixed effects due to the suppression of the gluon PDF.


\section{Conclusions}
\label{sec:Conclusions}
In this paper, we have considered the combined \qcdqed~transverse-momentum resummation formalism~\cite{Cieri:2018sfk,Autieri:2023xme,Buonocore:2024xmy} for the neutral current Drell-Yan (DY) process. Specifically, we dealt with the mixed \qcdqed~logarithmically-enhanced contributions, starting from the known $q_T$-subtraction coefficients of Ref.~\cite{Cieri:2020ikq} and the mixed renormalization group equations (RGE) for the QCD and QED coupling~\cite{Cieri:2018sfk, Billis:2019evv}. 

Here, we present for the first time all the required ingredients to consistently and simultaneously resum logarithmically enhanced contributions originating from QCD and QED emissions up to NNLL accuracy. In particular, we have obtained the resummed form factor including up to mixed \qcdqed~NNLL and pure QED NNLL effects for initial-state radiation. Although pure QED contributions were directly obtained by Abelianization ~\cite{deFlorian:2018wcj,AH:2019pyp,Cieri:2020ikq} of QCD counterparts \cite{Bozzi:2005wk}, mixed \qcdqed~terms were computed from first principles. Specifically, we relied on the integral representation of the Sudakov form factor, and the solution of RGE equations for QCD and QED coupling at different perturbative orders in a simultaneous \qcdqed~expansion. 

Then we applied the formalism to study the phenomenology at LHC (13 TeV) and Tevatron (1.96 TeV). For this purpose, we implemented the mixed higher-order resummation coefficients in the code \texttt{DYTURBO}~\cite{Camarda:2019zyx,Camarda:2021ict,Camarda:2021jsw,Camarda:2023dqn}. We studied the $q_T$ spectrum for $Z$/$\gamma^*$ production, comparing LL QED, mixed NLL \qcdqed~and mixed NNLL \qcdqed~resummation effects with respect to pure QCD corrections up to N$^3$LL. In all the cases, we found that NLL/NNLL \qcdqed~are smaller than LL QED corrections and they tend to stabilize for increasing $q_T$. Regarding the error bands induced by scale variations, we found that they are mainly driven by pure QCD uncertainties, and higher-order mixed \qcdqed~terms are more stable than LL QED ones. 

Besides this, the suppression of gluon-initiated processes at Tevatron leads, as expected, to an enhancement of pure QED and mixed \qcdqed~contributions w.r.t. LHC simulations. 

Also, we considered the corrections induced in the invariant mass distribution. In this case, we found that LL QED effects reduce up to $0.2 \, \%$ the distribution around the peak, and mixed NLL and NNLL go in the opposite direction: they increase the distribution approximately by $0.1 \, \%$ and $0.2 \, \%$, respectively. A similar behavior was found when considering the rapidity distribution (both at LHC and Tevatron), although the corrections were ${\cal O}(0.1 \, \%)$.

In conclusion, we have shown the impact of adding higher-order mixed \qcdqed ~\qt-resummation effects for $Z / \gamma*$ production to obtain more stable results, which is crucial in the context of the current high-precision physics program at colliders.


\section*{Acknowledgements}
We gratefully acknowledge Prasanna Dhani and Germ\'an Rodrigo for useful discussions. LC is supported by the Generalitat Valenciana (Spain) through the plan GenT program (CIDEGENT/2020/011) and his work is supported by the Spanish Government (Agencia Estatal de Investigación) and ERDF funds from European Commission (Grant no. PID2020-114473GB-I00 funded by MCIN/AEI/10.13039/501100011033). GS was partially supported by H2020-MSCA-COFUND USAL4EXCELLENCE-PROOPI-391 project under grant
agreement No. 101034371. GF is partially supported by the Italian Ministero dell'Universitá e Ricerca (MUR) through the research grant 20229KEFAM (PRIN2022, Next Generation EU, CUP H53D23000980006)


\appendix

\section{Effect of fixed QED coupling}
\label{app:Fixed}
The formalism presented in this article explicitly includes running effect in the QED coupling, in order to achieve a consistent treatment of gluon and photon radiation. However, we are aware that previous studies performed within the HEP community have been based on other approximations. For this reason, in this work, we also investigated the phenomenological impact of keeping the electromagnetic coupling fixed, i.e. $\alpha(\mu^2)=\alpha(m_Z^2)$ for all energy scales. 

\begin{figure}[htbp]
    \includegraphics[width=0.47\linewidth]{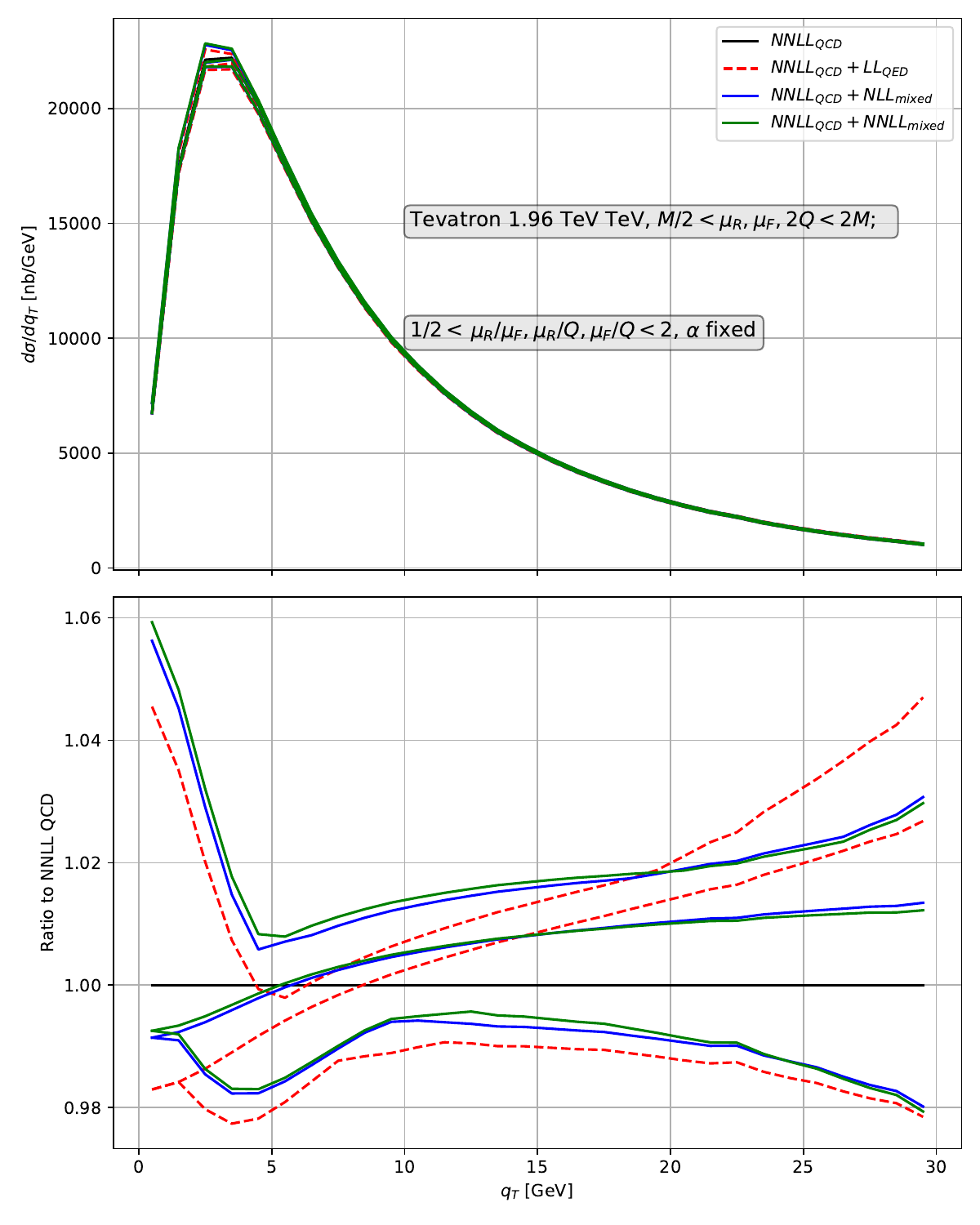}
     \includegraphics[width=0.47\linewidth]{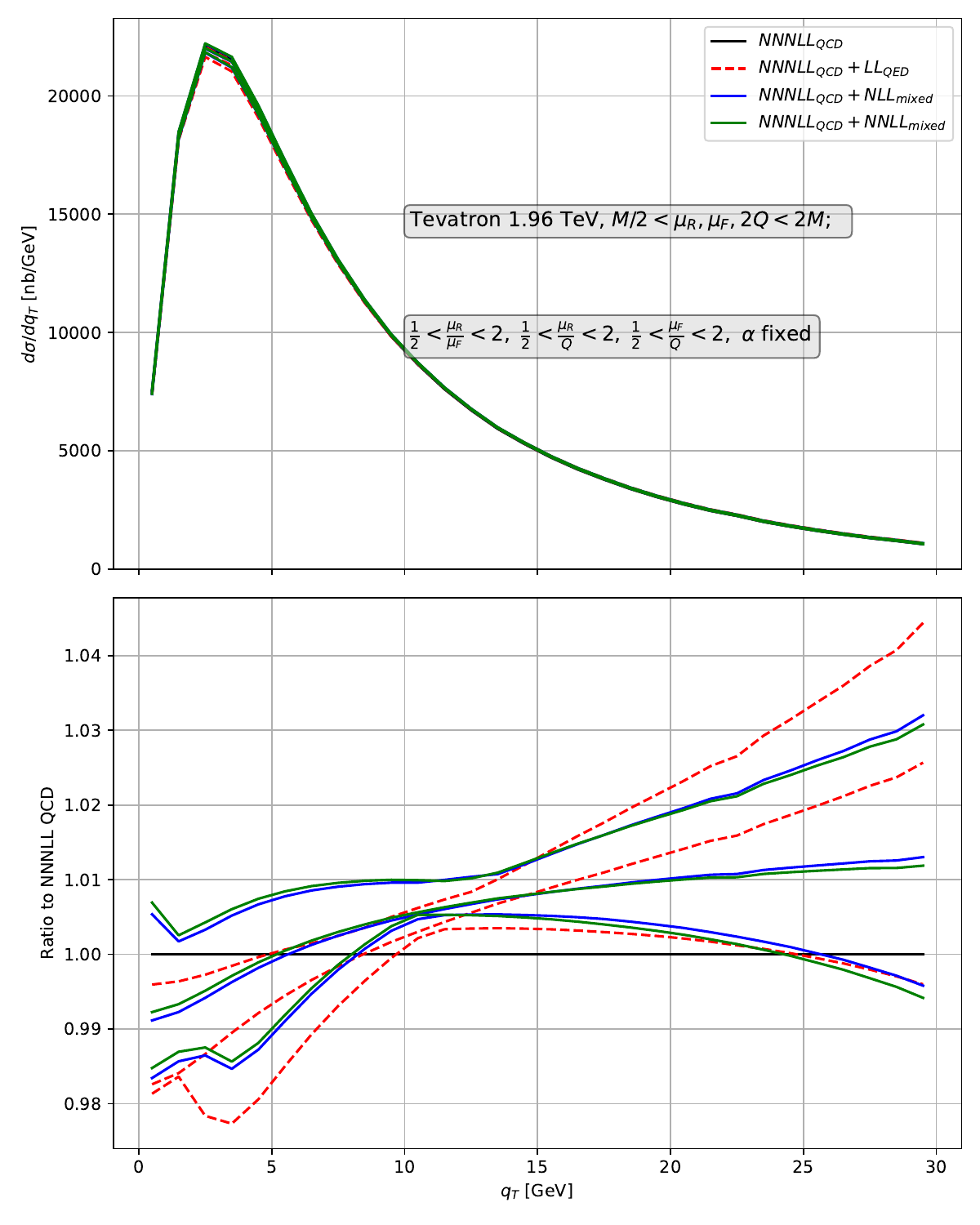}
    \caption{$Z/\gamma^*$ production at the Tevatron ($\sqrt{s} = 1.96 \,\text{TeV}$) at NNLL (left) and N$^3$LL (right) in QCD, including mixed \qcdqed~effects up to NNLL accuracy with $\alpha$ fixed. The lower panels show the ratio with respect to the pure QCD prediction, highlighting the impact of QED and mixed corrections. The uncertainty bands are obtained via scale variation, following the constraints indicated in the upper panels.}
    \label{fig:TevaNNNLLandQED_alphafix}
\end{figure}

In Fig.~\ref{fig:TevaNNNLLandQED_alphafix}, we present the $q_T$ distribution for $Z/\gamma^*$ production at $\sqrt{s} = 1.96 \,\text{TeV}$ for Tevatron, including up to N$^3$LL QCD and up to mixed NNLL \qcdqed~effects. On top of QCD prediction at NNLL, we observe that QED effects at LL accuracy range from approximately $-2\%$ to $+2\%$ in the considered $q_T$ region. These effects tend to harden the spectrum, shifting the peak toward higher $q_T$ values. The picture changes when including higher-order corrections: NLL and NNLL contributions continue to harden the spectrum, but with a reduced impact of about $\pm 1\%$. The mixed NNLL effects help to stabilize the NLL contribution.

Regarding the uncertainty bands, they are obtained by simultaneously varying the QCD and QED renormalization, factorization, and resummation scales. A good overlap of the bands is observed for $q_T \gtrsim 15\,\text{GeV}$, indicating good convergence of the mixed perturbative expansion. The widths of the uncertainty bands are ${\cal O}(5\text{--}7\%)$ at LL accuracy, and about the same ${\cal O}(5\text{--}7\%)$ at mixed NLL and NNLL.

Considering N$^3$LL predictions in QCD, we first observe a substantial reduction of the uncertainty bands, reflecting the higher precision of the QCD calculation. The qualitative behavior of the mixed corrections remains very similar to the previous case, with the LL component still producing the most pronounced hardening of the spectrum. QED effects are of the order $\pm 2 \%$ at LL and $\pm 1 \%$ at NLL and mixed NNLL. An overlap of the bands is observed for $q_T \gtrsim 15 \,\text{GeV}$, signaling again good convergence of the perturbative series. Scale variation bands are ${\cal O}(2\text{--}5\%)$ at LL and ${\cal O}(2\text{--}3\%)$ at mixed NLL and NNLL, respectively.

\bibliographystyle{JHEP}

\begin{thebibliography}{10}

\bibitem{Evans:2008zzb}
L.~Evans and P.~Bryant, \emph{{LHC Machine}}, \href{http://dx.doi.org/10.1088/1748-0221/3/08/S08001}{\emph{JINST} {\bf 3} (2008) S08001}.

\bibitem{Drell:1970wh}
S.~D. Drell and T.-M. Yan, \emph{{Massive Lepton Pair Production in Hadron-Hadron Collisions at High-Energies}}, \href{http://dx.doi.org/10.1103/PhysRevLett.25.316}{\emph{Phys. Rev. Lett.} {\bf 25} (1970) 316--320}.

\bibitem{Christenson:1970um}
J.~H. Christenson, G.~S. Hicks, L.~M. Lederman, P.~J. Limon, B.~G. Pope and E.~Zavattini, \emph{{Observation of massive muon pairs in hadron collisions}}, \href{http://dx.doi.org/10.1103/PhysRevLett.25.1523}{\emph{Phys. Rev. Lett.} {\bf 25} (1970) 1523--1526}.

\bibitem{Accomando:2017scx}
E.~Accomando, J.~Fiaschi, F.~Hautmann and S.~Moretti, \emph{{Constraining Parton Distribution Functions from Neutral Current Drell-Yan Measurements}}, \href{http://dx.doi.org/10.1103/PhysRevD.98.013003}{\emph{Phys. Rev. D} {\bf 98} (2018) 013003}, [\href{http://arxiv.org/abs/1712.06318}{{\tt 1712.06318}}].

\bibitem{CMS:2024ony}
{\scshape CMS} collaboration, A.~Hayrapetyan et~al., \emph{{Measurement of the Drell--Yan forward-backward asymmetry and of the effective leptonic weak mixing angle in proton-proton collisions at $\sqrt{s}$ = 13 TeV}},  \href{http://arxiv.org/abs/2408.07622}{{\tt 2408.07622}}.

\bibitem{CMS:2011utm}
{\scshape CMS} collaboration, S.~Chatrchyan et~al., \emph{{Measurement of the weak mixing angle with the Drell-Yan process in proton-proton collisions at the LHC}}, \href{http://dx.doi.org/10.1103/PhysRevD.84.112002}{\emph{Phys. Rev. D} {\bf 84} (2011) 112002}, [\href{http://arxiv.org/abs/1110.2682}{{\tt 1110.2682}}].

\bibitem{CDF:2013dpa}
{\scshape CDF, D0} collaboration, T.~A. Aaltonen et~al., \emph{{Combination of CDF and D0 $W$-Boson Mass Measurements}}, \href{http://dx.doi.org/10.1103/PhysRevD.88.052018}{\emph{Phys. Rev. D} {\bf 88} (2013) 052018}, [\href{http://arxiv.org/abs/1307.7627}{{\tt 1307.7627}}].

\bibitem{ATLAS:2017rzl}
{\scshape ATLAS} collaboration, M.~Aaboud et~al., \emph{{Measurement of the $W$-boson mass in pp collisions at $\sqrt{s}=7$ TeV with the ATLAS detector}}, \href{http://dx.doi.org/10.1140/epjc/s10052-017-5475-4}{\emph{Eur. Phys. J. C} {\bf 78} (2018) 110}, [\href{http://arxiv.org/abs/1701.07240}{{\tt 1701.07240}}].

\bibitem{CDF:2022hxs}
{\scshape CDF} collaboration, T.~Aaltonen et~al., \emph{{High-precision measurement of the $W$ boson mass with the CDF II detector}}, \href{http://dx.doi.org/10.1126/science.abk1781}{\emph{Science} {\bf 376} (2022) 170--176}.

\bibitem{ATLAS:2023fsi}
{\scshape ATLAS} collaboration, \emph{{Improved W boson Mass Measurement using 7 TeV Proton-Proton Collisions with the ATLAS Detector}}, .

\bibitem{Camarda:2016twt}
S.~Camarda, J.~Cuth and M.~Schott, \emph{{Determination of the muonic branching ratio of the W boson and its total width via cross-section measurements at the Tevatron and LHC}}, \href{http://dx.doi.org/10.1140/epjc/s10052-016-4461-6}{\emph{Eur. Phys. J. C} {\bf 76} (2016) 613}, [\href{http://arxiv.org/abs/1607.05084}{{\tt 1607.05084}}].

\bibitem{Camarda:2022qdg}
S.~Camarda, G.~Ferrera and M.~Schott, \emph{{Determination of the strong-coupling constant from the Z-boson transverse-momentum distribution}}, \href{http://dx.doi.org/10.1140/epjc/s10052-023-12373-2}{\emph{Eur. Phys. J. C} {\bf 84} (2024) 39}, [\href{http://arxiv.org/abs/2203.05394}{{\tt 2203.05394}}].

\bibitem{ATLAS:2023lhg}
{\scshape ATLAS} collaboration, G.~Aad et~al., \emph{{A precise determination of the strong-coupling constant from the recoil of $Z$ bosons with the ATLAS experiment at $\sqrt{s} = 8$ TeV}},  \href{http://arxiv.org/abs/2309.12986}{{\tt 2309.12986}}.

\bibitem{Rottoli:2023xdc}
L.~Rottoli, P.~Torrielli and A.~Vicini, \emph{{Determination of the W-boson mass at hadron colliders}}, \href{http://dx.doi.org/10.1140/epjc/s10052-023-12128-z}{\emph{Eur. Phys. J. C} {\bf 83} (2023) 948}, [\href{http://arxiv.org/abs/2301.04059}{{\tt 2301.04059}}].

\bibitem{CarloniCalame:2016ouw}
C.~M. Carloni~Calame, M.~Chiesa, H.~Martinez, G.~Montagna, O.~Nicrosini, F.~Piccinini et~al., \emph{{Precision Measurement of the W-Boson Mass: Theoretical Contributions and Uncertainties}}, \href{http://dx.doi.org/10.1103/PhysRevD.96.093005}{\emph{Phys. Rev. D} {\bf 96} (2017) 093005}, [\href{http://arxiv.org/abs/1612.02841}{{\tt 1612.02841}}].

\bibitem{Boughezal:2015dva}
R.~Boughezal, C.~Focke, X.~Liu and F.~Petriello, \emph{{$W$-boson production in association with a jet at next-to-next-to-leading order in perturbative QCD}}, \href{http://dx.doi.org/10.1103/PhysRevLett.115.062002}{\emph{Phys. Rev. Lett.} {\bf 115} (2015) 062002}, [\href{http://arxiv.org/abs/1504.02131}{{\tt 1504.02131}}].

\bibitem{Boughezal:2015ded}
R.~Boughezal, J.~M. Campbell, R.~K. Ellis, C.~Focke, W.~T. Giele, X.~Liu et~al., \emph{{Z-boson production in association with a jet at next-to-next-to-leading order in perturbative QCD}}, \href{http://dx.doi.org/10.1103/PhysRevLett.116.152001}{\emph{Phys. Rev. Lett.} {\bf 116} (2016) 152001}, [\href{http://arxiv.org/abs/1512.01291}{{\tt 1512.01291}}].

\bibitem{Gehrmann-DeRidder:2016cdi}
A.~Gehrmann-De~Ridder, T.~Gehrmann, E.~W.~N. Glover, A.~Huss and T.~A. Morgan, \emph{{The NNLO QCD corrections to Z boson production at large transverse momentum}}, \href{http://dx.doi.org/10.1007/JHEP07(2016)133}{\emph{JHEP} {\bf 07} (2016) 133}, [\href{http://arxiv.org/abs/1605.04295}{{\tt 1605.04295}}].

\bibitem{Gehrmann-DeRidder:2017mvr}
A.~Gehrmann-De~Ridder, T.~Gehrmann, E.~W.~N. Glover, A.~Huss and D.~M. Walker, \emph{{Next-to-Next-to-Leading-Order QCD Corrections to the Transverse Momentum Distribution of Weak Gauge Bosons}}, \href{http://dx.doi.org/10.1103/PhysRevLett.120.122001}{\emph{Phys. Rev. Lett.} {\bf 120} (2018) 122001}, [\href{http://arxiv.org/abs/1712.07543}{{\tt 1712.07543}}].

\bibitem{Chen:2022cgv}
X.~Chen, T.~Gehrmann, E.~W.~N. Glover, A.~Huss, P.~F. Monni, E.~Re et~al., \emph{{Third-Order Fiducial Predictions for Drell-Yan Production at the LHC}}, \href{http://dx.doi.org/10.1103/PhysRevLett.128.252001}{\emph{Phys. Rev. Lett.} {\bf 128} (2022) 252001}, [\href{http://arxiv.org/abs/2203.01565}{{\tt 2203.01565}}].

\bibitem{Alioli:2025hpa}
S.~Alioli, G.~Billis, A.~Broggio and G.~Stagnitto, \emph{{NNLO predictions with nonlocal subtractions and fiducial power corrections in GENEVA}}, \href{http://dx.doi.org/10.1007/JHEP01(2026)065}{\emph{JHEP} {\bf 01} (2026) 065}, [\href{http://arxiv.org/abs/2504.11357}{{\tt 2504.11357}}].

\bibitem{CMS:2019raw}
{\scshape CMS} collaboration, A.~M. Sirunyan et~al., \emph{{Measurements of differential Z boson production cross sections in proton-proton collisions at $ \sqrt{s} $ = 13 TeV}}, \href{http://dx.doi.org/10.1007/JHEP12(2019)061}{\emph{JHEP} {\bf 12} (2019) 061}, [\href{http://arxiv.org/abs/1909.04133}{{\tt 1909.04133}}].

\bibitem{ATLAS:2025hhn}
{\scshape ATLAS} collaboration, G.~Aad et~al., \emph{{Measurement of double-differential charged-current Drell-Yan cross-sections at high transverse masses in $pp$ collisions at $\sqrt{s} =$ 13 TeV with the ATLAS detector}},  \href{http://arxiv.org/abs/2502.21088}{{\tt 2502.21088}}.

\bibitem{ATLAS:2024nrd}
{\scshape ATLAS} collaboration, G.~Aad et~al., \emph{{Precise measurements of W- and Z-boson transverse momentum spectra with the ATLAS detector using pp collisions at $\sqrt{s} = 5.02$ TeV and 13 TeV}}, \href{http://dx.doi.org/10.1140/epjc/s10052-024-13414-0}{\emph{Eur. Phys. J. C} {\bf 84} (2024) 1126}, [\href{http://arxiv.org/abs/2404.06204}{{\tt 2404.06204}}].

\bibitem{ATLAS:2024irg}
{\scshape ATLAS} collaboration, G.~Aad et~al., \emph{{Measurement of vector boson production cross sections and their ratios using pp collisions at s=13.6 TeV with the ATLAS detector}}, \href{http://dx.doi.org/10.1016/j.physletb.2024.138725}{\emph{Phys. Lett. B} {\bf 854} (2024) 138725}, [\href{http://arxiv.org/abs/2403.12902}{{\tt 2403.12902}}].

\bibitem{CMS:2024myi}
{\scshape CMS} collaboration, A.~Hayrapetyan et~al., \emph{{Measurement of the inclusive cross sections for W and Z boson production in proton-proton collisions at $\sqrt{s}$ = 5.02 and 13 TeV}},  \href{http://arxiv.org/abs/2408.03744}{{\tt 2408.03744}}.

\bibitem{LHCb:2016fbk}
{\scshape LHCb} collaboration, R.~Aaij et~al., \emph{{Measurement of the forward Z boson production cross-section in pp collisions at $\sqrt{s} = 13$ TeV}}, \href{http://dx.doi.org/10.1007/JHEP09(2016)136}{\emph{JHEP} {\bf 09} (2016) 136}, [\href{http://arxiv.org/abs/1607.06495}{{\tt 1607.06495}}].

\bibitem{CarloniCalame:2006zq}
C.~M. Carloni~Calame, G.~Montagna, O.~Nicrosini and A.~Vicini, \emph{{Precision electroweak calculation of the charged current Drell-Yan process}}, \href{http://dx.doi.org/10.1088/1126-6708/2006/12/016}{\emph{JHEP} {\bf 12} (2006) 016}, [\href{http://arxiv.org/abs/hep-ph/0609170}{{\tt hep-ph/0609170}}].

\bibitem{CarloniCalame:2007cd}
C.~M. Carloni~Calame, G.~Montagna, O.~Nicrosini and A.~Vicini, \emph{{Precision electroweak calculation of the production of a high transverse-momentum lepton pair at hadron colliders}}, \href{http://dx.doi.org/10.1088/1126-6708/2007/10/109}{\emph{JHEP} {\bf 10} (2007) 109}, [\href{http://arxiv.org/abs/0710.1722}{{\tt 0710.1722}}].

\bibitem{Bonciani:2021zzf}
R.~Bonciani, L.~Buonocore, M.~Grazzini, S.~Kallweit, N.~Rana, F.~Tramontano et~al., \emph{{Mixed Strong-Electroweak Corrections to the Drell-Yan Process}}, \href{http://dx.doi.org/10.1103/PhysRevLett.128.012002}{\emph{Phys. Rev. Lett.} {\bf 128} (2022) 012002}, [\href{http://arxiv.org/abs/2106.11953}{{\tt 2106.11953}}].

\bibitem{Armadillo:2022bgm}
T.~Armadillo, R.~Bonciani, S.~Devoto, N.~Rana and A.~Vicini, \emph{{Two-loop mixed QCD-EW corrections to neutral current Drell-Yan}}, \href{http://dx.doi.org/10.1007/JHEP05(2022)072}{\emph{JHEP} {\bf 05} (2022) 072}, [\href{http://arxiv.org/abs/2201.01754}{{\tt 2201.01754}}].

\bibitem{Armadillo:2022ugh}
T.~Armadillo, R.~Bonciani, S.~Devoto, N.~Rana and A.~Vicini, \emph{{Evaluation of Feynman integrals with arbitrary complex masses via series expansions}}, \href{http://dx.doi.org/10.1016/j.cpc.2022.108545}{\emph{Comput. Phys. Commun.} {\bf 282} (2023) 108545}, [\href{http://arxiv.org/abs/2205.03345}{{\tt 2205.03345}}].

\bibitem{Buccioni:2022kgy}
F.~Buccioni, F.~Caola, H.~A. Chawdhry, F.~Devoto, M.~Heller, A.~von Manteuffel et~al., \emph{{Mixed QCD-electroweak corrections to dilepton production at the LHC in the high invariant mass region}}, \href{http://dx.doi.org/10.1007/JHEP06(2022)022}{\emph{JHEP} {\bf 06} (2022) 022}, [\href{http://arxiv.org/abs/2203.11237}{{\tt 2203.11237}}].

\bibitem{Buonocore:2021rxx}
L.~Buonocore, M.~Grazzini, S.~Kallweit, C.~Savoini and F.~Tramontano, \emph{{Mixed QCD-EW corrections to $\boldsymbol{pp\!\to\!\ell\nu_\ell\!+\!X}$ at the LHC}}, \href{http://dx.doi.org/10.1103/PhysRevD.103.114012}{\emph{Phys. Rev. D} {\bf 103} (2021) 114012}, [\href{http://arxiv.org/abs/2102.12539}{{\tt 2102.12539}}].

\bibitem{Parisi:1979se}
G.~Parisi and R.~Petronzio, \emph{{Small Transverse Momentum Distributions in Hard Processes}}, \href{http://dx.doi.org/10.1016/0550-3213(79)90040-3}{\emph{Nucl. Phys. B} {\bf 154} (1979) 427--440}.

\bibitem{Kodaira:1981nh}
J.~Kodaira and L.~Trentadue, \emph{{Summing Soft Emission in QCD}}, \href{http://dx.doi.org/10.1016/0370-2693(82)90907-8}{\emph{Phys. Lett. B} {\bf 112} (1982) 66}.

\bibitem{Altarelli:1984pt}
G.~Altarelli, R.~K. Ellis, M.~Greco and G.~Martinelli, \emph{{Vector Boson Production at Colliders: A Theoretical Reappraisal}}, \href{http://dx.doi.org/10.1016/0550-3213(84)90112-3}{\emph{Nucl. Phys. B} {\bf 246} (1984) 12--44}.

\bibitem{Collins:1984kg}
J.~C. Collins, D.~E. Soper and G.~F. Sterman, \emph{{Transverse Momentum Distribution in Drell-Yan Pair and W and Z Boson Production}}, \href{http://dx.doi.org/10.1016/0550-3213(85)90479-1}{\emph{Nucl. Phys. B} {\bf 250} (1985) 199--224}.

\bibitem{Catani:2000vq}
S.~Catani, D.~de~Florian and M.~Grazzini, \emph{{Universality of nonleading logarithmic contributions in transverse momentum distributions}}, \href{http://dx.doi.org/10.1016/S0550-3213(00)00617-9}{\emph{Nucl. Phys. B} {\bf 596} (2001) 299--312}, [\href{http://arxiv.org/abs/hep-ph/0008184}{{\tt hep-ph/0008184}}].

\bibitem{Camarda:2021ict}
S.~Camarda, L.~Cieri and G.~Ferrera, \emph{{Drell\textendash{}Yan lepton-pair production: qT resummation at N3LL accuracy and fiducial cross sections at N3LO}}, \href{http://dx.doi.org/10.1103/PhysRevD.104.L111503}{\emph{Phys. Rev. D} {\bf 104} (2021) L111503}, [\href{http://arxiv.org/abs/2103.04974}{{\tt 2103.04974}}].

\bibitem{Bizon:2018foh}
W.~Bizo\'n, X.~Chen, A.~Gehrmann-De~Ridder, T.~Gehrmann, N.~Glover, A.~Huss et~al., \emph{{Fiducial distributions in Higgs and Drell-Yan production at N$^{3}$LL+NNLO}}, \href{http://dx.doi.org/10.1007/JHEP12(2018)132}{\emph{JHEP} {\bf 12} (2018) 132}, [\href{http://arxiv.org/abs/1805.05916}{{\tt 1805.05916}}].

\bibitem{Bizon:2019zgf}
W.~Bizon, A.~Gehrmann-De~Ridder, T.~Gehrmann, N.~Glover, A.~Huss, P.~F. Monni et~al., \emph{{The transverse momentum spectrum of weak gauge bosons at N ${}^3$ LL + NNLO}}, \href{http://dx.doi.org/10.1140/epjc/s10052-019-7324-0}{\emph{Eur. Phys. J. C} {\bf 79} (2019) 868}, [\href{http://arxiv.org/abs/1905.05171}{{\tt 1905.05171}}].

\bibitem{Bacchetta:2019sam}
A.~Bacchetta, V.~Bertone, C.~Bissolotti, G.~Bozzi, F.~Delcarro, F.~Piacenza et~al., \emph{{Transverse-momentum-dependent parton distributions up to N$^{3}$LL from Drell-Yan data}}, \href{http://dx.doi.org/10.1007/JHEP07(2020)117}{\emph{JHEP} {\bf 07} (2020) 117}, [\href{http://arxiv.org/abs/1912.07550}{{\tt 1912.07550}}].

\bibitem{Ebert:2020dfc}
M.~A. Ebert, J.~K.~L. Michel, I.~W. Stewart and F.~J. Tackmann, \emph{{Drell-Yan $q_{T}$ resummation of fiducial power corrections at N$^{3}$LL}}, \href{http://dx.doi.org/10.1007/JHEP04(2021)102}{\emph{JHEP} {\bf 04} (2021) 102}, [\href{http://arxiv.org/abs/2006.11382}{{\tt 2006.11382}}].

\bibitem{Becher:2020ugp}
T.~Becher and T.~Neumann, \emph{{Fiducial $q_T$ resummation of color-singlet processes at N$^3$LL+NNLO}}, \href{http://dx.doi.org/10.1007/JHEP03(2021)199}{\emph{JHEP} {\bf 03} (2021) 199}, [\href{http://arxiv.org/abs/2009.11437}{{\tt 2009.11437}}].

\bibitem{Re:2021con}
E.~Re, L.~Rottoli and P.~Torrielli, \emph{{Fiducial Higgs and Drell-Yan distributions at N$^3$LL$^\prime$+NNLO with RadISH}},  \href{http://arxiv.org/abs/2104.07509}{{\tt 2104.07509}}.

\bibitem{Ju:2021lah}
W.-L. Ju and M.~Sch\"onherr, \emph{{The q$_{T}$ and \ensuremath{\Delta}\ensuremath{\phi} spectra in W and Z production at the LHC at N$^{3}$LL'+N$^{2}$LO}}, \href{http://dx.doi.org/10.1007/JHEP10(2021)088}{\emph{JHEP} {\bf 10} (2021) 088}, [\href{http://arxiv.org/abs/2106.11260}{{\tt 2106.11260}}].

\bibitem{Campbell:2022uzw}
J.~M. Campbell, R.~K. Ellis, T.~Neumann and S.~Seth, \emph{{Transverse momentum resummation at N$^{3}$LL+NNLO for diboson processes}}, \href{http://dx.doi.org/10.1007/JHEP03(2023)080}{\emph{JHEP} {\bf 03} (2023) 080}, [\href{http://arxiv.org/abs/2210.10724}{{\tt 2210.10724}}].

\bibitem{Campbell:2023cha}
J.~M. Campbell, R.~K. Ellis, T.~Neumann and S.~Seth, \emph{{Jet-veto resummation at N$^{3}$LL$_{p}$ + NNLO in boson production processes}}, \href{http://dx.doi.org/10.1007/JHEP04(2023)106}{\emph{JHEP} {\bf 04} (2023) 106}, [\href{http://arxiv.org/abs/2301.11768}{{\tt 2301.11768}}].

\bibitem{Isaacson:2023iui}
J.~Isaacson, Y.~Fu and C.~P. Yuan, \emph{{Improving resbos for the precision needs of the LHC}}, \href{http://dx.doi.org/10.1103/PhysRevD.110.073002}{\emph{Phys. Rev. D} {\bf 110} (2024) 073002}, [\href{http://arxiv.org/abs/2311.09916}{{\tt 2311.09916}}].

\bibitem{Bacchetta:2022awv}
{\scshape MAP (Multi-dimensional Analyses of Partonic distributions)} collaboration, A.~Bacchetta, V.~Bertone, C.~Bissolotti, G.~Bozzi, M.~Cerutti, F.~Piacenza et~al., \emph{{Unpolarized transverse momentum distributions from a global fit of Drell-Yan and semi-inclusive deep-inelastic scattering data}}, \href{http://dx.doi.org/10.1007/JHEP10(2022)127}{\emph{JHEP} {\bf 10} (2022) 127}, [\href{http://arxiv.org/abs/2206.07598}{{\tt 2206.07598}}].

\bibitem{Bacchetta:2024qre}
{\scshape MAP (Multi-dimensional Analyses of Partonic distributions)} collaboration, A.~Bacchetta, V.~Bertone, C.~Bissolotti, G.~Bozzi, M.~Cerutti, F.~Delcarro et~al., \emph{{Flavor dependence of unpolarized quark transverse momentum distributions from a global fit}}, \href{http://dx.doi.org/10.1007/JHEP08(2024)232}{\emph{JHEP} {\bf 08} (2024) 232}, [\href{http://arxiv.org/abs/2405.13833}{{\tt 2405.13833}}].

\bibitem{Bacchetta:2025ara}
{\scshape MAP (Multi-dimensional Analyses of Partonic distributions)} collaboration, A.~Bacchetta, V.~Bertone, C.~Bissolotti, M.~Cerutti, M.~Radici, S.~Rodini et~al., \emph{{Neural-Network Extraction of Unpolarized Transverse-Momentum-Dependent Distributions}}, \href{http://dx.doi.org/10.1103/csc2-bj91}{\emph{Phys. Rev. Lett.} {\bf 135} (2025) 021904}, [\href{http://arxiv.org/abs/2502.04166}{{\tt 2502.04166}}].

\bibitem{Camarda:2025lbt}
S.~Camarda, G.~Ferrera and L.~Rossi, \emph{{Drell--Yan lepton pair production at low invariant masses: transverse-momentum resummation and non-perturbative effects in QCD}},  \href{http://arxiv.org/abs/2508.06201}{{\tt 2508.06201}}.

\bibitem{Neumann:2022lft}
T.~Neumann and J.~Campbell, \emph{{Fiducial Drell-Yan production at the LHC improved by transverse-momentum resummation at N4LLp+N3LO}}, \href{http://dx.doi.org/10.1103/PhysRevD.107.L011506}{\emph{Phys. Rev. D} {\bf 107} (2023) L011506}, [\href{http://arxiv.org/abs/2207.07056}{{\tt 2207.07056}}].

\bibitem{Camarda:2023dqn}
S.~Camarda, L.~Cieri and G.~Ferrera, \emph{{Drell\textendash{}Yan lepton-pair production: qT resummation at N4LL accuracy}}, \href{http://dx.doi.org/10.1016/j.physletb.2023.138125}{\emph{Phys. Lett. B} {\bf 845} (2023) 138125}, [\href{http://arxiv.org/abs/2303.12781}{{\tt 2303.12781}}].

\bibitem{Moos:2025sal}
V.~Moos, I.~Scimemi, A.~Vladimirov and P.~Zurita, \emph{{Determination of unpolarized TMD distributions from the fit of Drell-Yan and SIDIS data at N$^4$LL}},  \href{http://arxiv.org/abs/2503.11201}{{\tt 2503.11201}}.

\bibitem{Billis:2024dqq}
G.~Billis, J.~K.~L. Michel and F.~J. Tackmann, \emph{{Drell-Yan transverse-momentum spectra at N$^{3}$LL$^{'}$ and approximate N$^{4}$LL with SCETlib}}, \href{http://dx.doi.org/10.1007/JHEP02(2025)170}{\emph{JHEP} {\bf 02} (2025) 170}, [\href{http://arxiv.org/abs/2411.16004}{{\tt 2411.16004}}].

\bibitem{Piloneta:2024aac}
S.~Piloneta and A.~Vladimirov, \emph{{Angular distributions of Drell-Yan leptons in the TMD factorization approach}}, \href{http://dx.doi.org/10.1007/JHEP12(2024)059}{\emph{JHEP} {\bf 12} (2024) 059}, [\href{http://arxiv.org/abs/2407.06277}{{\tt 2407.06277}}].

\bibitem{Cieri:2018sfk}
L.~Cieri, G.~Ferrera and G.~F.~R. Sborlini, \emph{{Combining QED and QCD transverse-momentum resummation for Z boson production at hadron colliders}}, \href{http://dx.doi.org/10.1007/JHEP08(2018)165}{\emph{JHEP} {\bf 08} (2018) 165}, [\href{http://arxiv.org/abs/1805.11948}{{\tt 1805.11948}}].

\bibitem{Autieri:2023xme}
A.~Autieri, L.~Cieri, G.~Ferrera and G.~F.~R. Sborlini, \emph{{Combining QED and QCD transverse-momentum resummation for W and Z boson production at hadron colliders}}, \href{http://dx.doi.org/10.1007/JHEP07(2023)104}{\emph{JHEP} {\bf 07} (2023) 104}, [\href{http://arxiv.org/abs/2302.05403}{{\tt 2302.05403}}].

\bibitem{Buonocore:2024xmy}
L.~Buonocore, L.~Rottoli and P.~Torrielli, \emph{{Resummation of combined QCD-electroweak effects in Drell Yan lepton-pair production}}, \href{http://dx.doi.org/10.1007/JHEP07(2024)193}{\emph{JHEP} {\bf 07} (2024) 193}, [\href{http://arxiv.org/abs/2404.15112}{{\tt 2404.15112}}].

\bibitem{Bozzi:2005wk}
G.~Bozzi, S.~Catani, D.~de~Florian and M.~Grazzini, \emph{{Transverse-momentum resummation and the spectrum of the Higgs boson at the LHC}}, \href{http://dx.doi.org/10.1016/j.nuclphysb.2005.12.022}{\emph{Nucl. Phys. B} {\bf 737} (2006) 73--120}, [\href{http://arxiv.org/abs/hep-ph/0508068}{{\tt hep-ph/0508068}}].

\bibitem{Collins:1989gx}
J.~C. Collins, D.~E. Soper and G.~F. Sterman, \emph{{Factorization of Hard Processes in QCD}}, \href{http://dx.doi.org/10.1142/9789814503266_0001}{\emph{Adv. Ser. Direct. High Energy Phys.} {\bf 5} (1989) 1--91}, [\href{http://arxiv.org/abs/hep-ph/0409313}{{\tt hep-ph/0409313}}].

\bibitem{Bozzi:2010xn}
G.~Bozzi, S.~Catani, G.~Ferrera, D.~de~Florian and M.~Grazzini, \emph{{Production of Drell-Yan lepton pairs in hadron collisions: Transverse-momentum resummation at next-to-next-to-leading logarithmic accuracy}}, \href{http://dx.doi.org/10.1016/j.physletb.2010.12.024}{\emph{Phys. Lett. B} {\bf 696} (2011) 207--213}, [\href{http://arxiv.org/abs/1007.2351}{{\tt 1007.2351}}].

\bibitem{Bizon:2017rah}
W.~Bizon, P.~F. Monni, E.~Re, L.~Rottoli and P.~Torrielli, \emph{{Momentum-space resummation for transverse observables and the Higgs p$_{\perp}$ at N$^{3}$LL+NNLO}}, \href{http://dx.doi.org/10.1007/JHEP02(2018)108}{\emph{JHEP} {\bf 02} (2018) 108}, [\href{http://arxiv.org/abs/1705.09127}{{\tt 1705.09127}}].

\bibitem{Catani:2012qa}
S.~Catani, L.~Cieri, D.~de~Florian, G.~Ferrera and M.~Grazzini, \emph{{Vector boson production at hadron colliders: hard-collinear coefficients at the NNLO}}, \href{http://dx.doi.org/10.1140/epjc/s10052-012-2195-7}{\emph{Eur. Phys. J. C} {\bf 72} (2012) 2195}, [\href{http://arxiv.org/abs/1209.0158}{{\tt 1209.0158}}].

\bibitem{Catani:2013tia}
S.~Catani, L.~Cieri, D.~de~Florian, G.~Ferrera and M.~Grazzini, \emph{{Universality of transverse-momentum resummation and hard factors at the NNLO}}, \href{http://dx.doi.org/10.1016/j.nuclphysb.2014.02.011}{\emph{Nucl. Phys. B} {\bf 881} (2014) 414--443}, [\href{http://arxiv.org/abs/1311.1654}{{\tt 1311.1654}}].

\bibitem{Gehrmann:2012ze}
T.~Gehrmann, T.~Lubbert and L.~L. Yang, \emph{{Transverse parton distribution functions at next-to-next-to-leading order: the quark-to-quark case}}, \href{http://dx.doi.org/10.1103/PhysRevLett.109.242003}{\emph{Phys. Rev. Lett.} {\bf 109} (2012) 242003}, [\href{http://arxiv.org/abs/1209.0682}{{\tt 1209.0682}}].

\bibitem{Luo:2019szz}
M.-x. Luo, T.-Z. Yang, H.~X. Zhu and Y.~J. Zhu, \emph{{Quark Transverse Parton Distribution at the Next-to-Next-to-Next-to-Leading Order}}, \href{http://dx.doi.org/10.1103/PhysRevLett.124.092001}{\emph{Phys. Rev. Lett.} {\bf 124} (2020) 092001}, [\href{http://arxiv.org/abs/1912.05778}{{\tt 1912.05778}}].

\bibitem{Ebert:2020yqt}
M.~A. Ebert, B.~Mistlberger and G.~Vita, \emph{{Transverse momentum dependent PDFs at N$^3$LO}}, \href{http://dx.doi.org/10.1007/JHEP09(2020)146}{\emph{JHEP} {\bf 09} (2020) 146}, [\href{http://arxiv.org/abs/2006.05329}{{\tt 2006.05329}}].

\bibitem{Billis:2019evv}
G.~Billis, F.~J. Tackmann and J.~Talbert, \emph{{Higher-Order Sudakov Resummation in Coupled Gauge Theories}}, \href{http://dx.doi.org/10.1007/JHEP03(2020)182}{\emph{JHEP} {\bf 03} (2020) 182}, [\href{http://arxiv.org/abs/1907.02971}{{\tt 1907.02971}}].

\bibitem{deFlorian:2016gvk}
D.~de~Florian, G.~F.~R. Sborlini and G.~Rodrigo, \emph{{Two-loop QED corrections to the Altarelli-Parisi splitting functions}}, \href{http://dx.doi.org/10.1007/JHEP10(2016)056}{\emph{JHEP} {\bf 10} (2016) 056}, [\href{http://arxiv.org/abs/1606.02887}{{\tt 1606.02887}}].

\bibitem{deFlorian:2015ujt}
D.~de~Florian, G.~F.~R. Sborlini and G.~Rodrigo, \emph{{QED corrections to the Altarelli{\textendash}Parisi splitting functions}}, \href{http://dx.doi.org/10.1140/epjc/s10052-016-4131-8}{\emph{Eur. Phys. J. C} {\bf 76} (2016) 282}, [\href{http://arxiv.org/abs/1512.00612}{{\tt 1512.00612}}].

\bibitem{deFlorian:2018wcj}
D.~de~Florian, M.~Der and I.~Fabre, \emph{{QCD$\oplus$QED NNLO corrections to Drell Yan production}}, \href{http://dx.doi.org/10.1103/PhysRevD.98.094008}{\emph{Phys. Rev. D} {\bf 98} (2018) 094008}, [\href{http://arxiv.org/abs/1805.12214}{{\tt 1805.12214}}].

\bibitem{Cieri:2020ikq}
L.~Cieri, D.~de~Florian, M.~Der and J.~Mazzitelli, \emph{{Mixed QCD\ensuremath{\otimes}QED corrections to exclusive Drell Yan production using the q$_{T}$ -subtraction method}}, \href{http://dx.doi.org/10.1007/JHEP09(2020)155}{\emph{JHEP} {\bf 09} (2020) 155}, [\href{http://arxiv.org/abs/2005.01315}{{\tt 2005.01315}}].

\bibitem{AH:2019pyp}
A.~H. Ajjath, P.~Mukherjee and V.~Ravindran, \emph{{Infrared structure of $SU(N)\times U(1)$ gauge theory to three loops}}, \href{http://dx.doi.org/10.1007/JHEP08(2020)156}{\emph{JHEP} {\bf 08} (2020) 156}, [\href{http://arxiv.org/abs/1912.13386}{{\tt 1912.13386}}].

\bibitem{Becher:2010tm}
T.~Becher and M.~Neubert, \emph{{Drell-Yan Production at Small $q_T$, Transverse Parton Distributions and the Collinear Anomaly}}, \href{http://dx.doi.org/10.1140/epjc/s10052-011-1665-7}{\emph{Eur. Phys. J. C} {\bf 71} (2011) 1665}, [\href{http://arxiv.org/abs/1007.4005}{{\tt 1007.4005}}].

\bibitem{Stadlbauer:2024jij}
M.~Stadlbauer and T.~Theil, \emph{{Infrared anomalous dimensions at three-loop in the SM from conserved currents}}, \href{http://dx.doi.org/10.1007/JHEP01(2025)050}{\emph{JHEP} {\bf 01} (2025) 050}, [\href{http://arxiv.org/abs/2409.15422}{{\tt 2409.15422}}].

\bibitem{ZENODO}
A.~Autieri, S.~Camarda, L.~Cieri, G.~Ferrera and G.~Sborlini, \emph{{Ancillary files for "Transverse-momentum resummation at mixed QCD-QED NNLL accuracy for Z boson production at hadron colliders"}},  2025.
\newblock 10.5281/zenodo.15189083.

\bibitem{Camarda:2019zyx}
S.~Camarda et~al., \emph{{DYTurbo: Fast predictions for Drell-Yan processes}}, \href{http://dx.doi.org/10.1140/epjc/s10052-020-7757-5}{\emph{Eur. Phys. J. C} {\bf 80} (2020) 251}, [\href{http://arxiv.org/abs/1910.07049}{{\tt 1910.07049}}].

\bibitem{Camarda:2021jsw}
S.~Camarda, L.~Cieri and G.~Ferrera, \emph{{Fiducial perturbative power corrections within the $\mathbf{q}_T$ subtraction formalism}}, \href{http://dx.doi.org/10.1140/epjc/s10052-022-10510-x}{\emph{Eur. Phys. J. C} {\bf 82} (2022) 575}, [\href{http://arxiv.org/abs/2111.14509}{{\tt 2111.14509}}].

\bibitem{NNPDF:2024djq}
{\scshape NNPDF} collaboration, R.~D. Ball et~al., \emph{{Photons in the proton: implications for the LHC}}, \href{http://dx.doi.org/10.1140/epjc/s10052-024-12731-8}{\emph{Eur. Phys. J. C} {\bf 84} (2024) 540}, [\href{http://arxiv.org/abs/2401.08749}{{\tt 2401.08749}}].

\end{thebibliography}

\providecommand{\href}[2]{#2}\begingroup\raggedright\endgroup


\end{document}